\newcommand{\cov}{COVID-19}
\newcommand{\etal}{{\em et al}}
\newcommand{\numcountries}{40}
\newcommand{\pmt}{PM$_{2.5}$}

\documentclass[10pt, a4paper, twocolumn]{article} 

\usepackage[english]{babel} 

\usepackage{microtype} 

\usepackage{amsmath,amsfonts,amsthm} 

\usepackage[svgnames]{xcolor} 

\usepackage[hang, small, labelfont=bf, up, textfont=it]{caption} 

\usepackage{booktabs} 

\usepackage{lastpage} 

\usepackage{graphicx} 

\usepackage{enumitem} 
\setlist{noitemsep} 

\usepackage{sectsty} 
\allsectionsfont{\usefont{OT1}{phv}{b}{n}} 

\usepackage{geometry} 

\usepackage[T1]{fontenc} 
\usepackage[utf8]{inputenc} 

\usepackage{XCharter} 

\usepackage{fancyhdr} 
\fancypagestyle{firstpage}{ 
	\fancyhf{}
}

\newcommand{\authorstyle}[1]{{\large\usefont{OT1}{phv}{b}{n}\color{Black}#1}} 

\newcommand{\institution}[1]{{\footnotesize\usefont{OT1}{phv}{m}{sl}\color{Black}#1}} 

\usepackage{titling} 

\newcommand{\HorRule}{\color{DarkGoldenrod}\rule{\linewidth}{1pt}} 

\pretitle{
	\vspace{-30pt} 
	\HorRule\vspace{10pt} 
	\fontsize{23}{33}\usefont{OT1}{phv}{b}{n}\selectfont 
	\color{DarkRed} 
}

\posttitle{\par\vskip 15pt} 

\preauthor{} 

\postauthor{ 
	\vspace{10pt} 
	\par\HorRule 
	\vspace{20pt} 
}

\usepackage{lettrine} 
\usepackage{fix-cm}	

\newcommand{\initial}[1]{ 
	\lettrine[lines=3,findent=4pt,nindent=0pt]{
		\color{DarkGoldenrod}
		{#1}
	}{}%
}

\usepackage{xstring} 

\newcommand{\lettrineabstract}[1]{
	\StrLeft{#1}{1}[\firstletter] 
	\initial{\firstletter}\textbf{\StrGobbleLeft{#1}{1}} 
}

\usepackage[autostyle=true]{csquotes} 

\usepackage{comment}

\title{Climate \& BCG: Effects on \cov\ Death Growth Rates} 

\author{
	\authorstyle{Chris Finlay \textsuperscript{1,2,3} and Bruce Bassett\textsuperscript{2,3}} 
	\newline\newline 
	\textsuperscript{1}\institution{Universidad Nacional Autónoma de México, Mexico City, Mexico}\\ 
	\textsuperscript{2}\institution{University of Texas at Austin, Texas, United States of America}\\ 
	\textsuperscript{3}\institution{\texttt{LaTeXTemplates.com}} 
}

\author{
	\authorstyle{Chris Finlay\textsuperscript{1,2,3,+} and Bruce A. Bassett  \textsuperscript{1,2,3,4,*}} 
	\newline\newline 
	\textsuperscript{1}\institution{South African Radio Astronomical Observatory, Observatory, Cape Town, 7295}\\
	\textsuperscript{2}\institution{Department of Maths and Applied Maths, University of Cape Town, Rondebosch, Cape Town, 7700}\\ 
	\textsuperscript{3}\institution{African Institute for Mathematical Sciences, Muizenburg, Cape Town, 7950, South Africa}\\ %
	\textsuperscript{4}\institution{South African Astronomical Observatory, Observatory, Cape Town, 7295}\\ 
	\textsuperscript{+}\institution{Email:cfinlay@ska.ac.za}	\textsuperscript{*}\institution{Email:bruce.a.bassett@gmail.com}
}

\date{\today} 

\begin{document}

\maketitle 

\thispagestyle{firstpage} 


\lettrineabstract{Multiple studies have suggested the spread of COVID-19 is affected by factors such as climate, BCG vaccinations, pollution and blood type. We perform a joint study of these factors using the death growth rates of 40 regions worldwide with both machine learning and Bayesian methods. We find weak,  non-significant ($< 3\sigma$) evidence for temperature and relative humidity as factors in the spread of \cov\ but little or no evidence for BCG vaccination prevalence or \pmt\ pollution. The only variable detected at a statistically significant level (>$3\sigma$) is the rate of positive \cov\ tests, with higher positive rates correlating with higher daily growth of deaths.}


\section{Introduction}

The \cov\ pandemic has triggered extensive efforts to predict the  severity of \cov\ to aid in decision making around  interventions such as lockdown and the closure of schools \footnote{https://www.imperial.ac.uk/media/imperial-college/medicine/sph/ide/gida-fellowships/Imperial-College-COVID19-NPI-modelling-16-03-2020.pdf} \cite{imperial2}. Regions  hit hard by the pandemic, such as Wuhan, Lombardy and New York, where doubling times of 2-3 days and high crude mortality rates stand in stark contrast to other countries only mildly affected such as Hong Kong, South Korea and New Zealand. 

Potential explanations for the apparent differences in the transmissivity (encoded by the time-dependent reproductive number, $R_t$) and lethality (encoded by the Infection Fatality Rate, IFR), currently fall into four broad categories. The first posits that differences are largely fictitious, driven by the heterogeneity of testing and reporting of cases and deaths; a known issue and one that we are particularly concerned with in this paper. 

The next three categories posit that the differences are primarily real and are driven by (1) cultural and policy factors (swift lockdown, efficient contact tracing and quarantining, use of masks, obedient populations or social structures that are naturally distant or isolated), (2) local environmental factors (such as temperature and humidity variations, population density, comorbid factors, vaccinations, vitamin D levels, blood type etc...) or (3) existence of multiple strains with different transmissibility or lethality \cite{strains}. Our primary interest lies in disentangling the first category (testing) from a subset of potential factors in the second category.

Finding the relative contributions of each of these four categories is key in understanding and optimally fighting the pandemic. The widely different testing capabilities between countries, particularly between the developed and developing world, imply that if not correctly treated, testing variability will create spurious evidence that can lead to false hope and sub-optimal interventions. 

In the wake of the spreading pandemic there have been a host of studies that have examined the possible impact of climate \cite{weather1}-\cite{weather12}, blood type \cite{blood1,blood2}, haplogroup \cite{haplogroup}, pollution \cite{pollution}-\cite{pollution5} and  BCG vaccination prevalence \cite{bcg1,bcg2,bcg_neg, bcg_neg2} on the spread of \cov. Our main conclusion is that testing issues are significant and the factors above are likely not the main causes of variability in growth rates of deaths worldwide. 

We note that both the basic reproductive number, $R_0$, and the IFR will be affected by the four categories above in general. However, since we do not currently have access to the true number of infections we cannot address the potential effect of environmental factors on the IFR. Similarly the Case Fatality Rate cannot be used for this purpose due to testing differences around the world. We therefore focus on their potential effect on $R_0$, which we quantify through the daily growth rate of deaths of countries around the world for which we have sufficiently good data. 

\section{Methodology}

To explore the potential impact of climate (e.g. through temperature, relative humidity and UV Index), BCG vaccination prevalence, blood type and pollution (\pmt) on \cov, one must think carefully about choice of both data and methodology. 

We choose to focus on deaths instead of the number of confirmed infections in the belief that these are significantly less affected by sampling and testing issues than infection numbers: the number of tests per 1000 population (Tests/1k) currently varies by more than three orders of magnitude across the world \footnote{http://worldometers.info/coronavirus/}, making infection numbers highly biased and correlated with confounding variables such as GDP and healthcare. In countries with limited testing capability, only the more severe patients are typically tested. 

Since these are also the patients most likely to die, it is likely that patients who die from \cov\ are more likely to be tested than typical \cov\ sufferers, who may be mostly asymptomatic. Without detailed knowledge of testing protocols in each country separating out the effects of the factors of interest (climate, BCG vaccination etc...) is extremely challenging. Although deaths are not immune to testing issues, with many missing deaths shown through excess mortality studies, especially in overwhelmed medical systems \cite{excess1, excess2}, we argue that this is likely to be much less of an issue in the first month after initial \cov\ deaths, which we focus on, since we are interested in $R_0$. 

The second key choice is whether to focus on absolute death counts or growth rates. Absolute death counts are also highly problematic. First, testing efficiencies may vary systematically from country to country. Second, the widely varying start dates of the pandemic in different countries are hard to deal with rigorously. As a result we focus on the daily growth rate of deaths, $G_c$, for country $c$ during the initial phase of the epidemic. The initial daily growth correlates with $R_{0c}$, the base reproductive number in each country; in simple models $R_{0c} = (1+G_c)^{\tau}$, where $\tau$ is the period in days for which a person is infectious on average\footnote{Hence  $G_c=0.3$ and $\tau = 5$ days yields $R_0 \simeq 3.7$.}.

Separating out the true causes of variation in $G_c$ or $R_{0c}$ is still a highly complex problem, and we discuss our approach in detail in the additional material in section (\ref{sec:additional}). Part of the complexity arises from the fact that $R_{0c}$ depends explicitly both on properties of the virus and on social factors (e.g. average number and nature of contacts), as well as any of the other potential factors we wish to study.  Hence, we expect there to be a large intrinsic variation in the growth of deaths from country to country, driven e.g. by the average number of people in a household, population density \cite{luis} etc...  that may correlate with, and hence lead spurious evidence for, potential factors such as climate, vaccination coverage, blood type and pollution. 

We model this complexity by allowing each country, $c$, to have its own unique base growth rate, denoted $G_{0c}$, that is estimated from its own data. This freedom allows the model to account for the myriad unmodelled factors specific to each country (population density, GDP, culture, health care quality etc...). However, we tie these base country growth rates, $G_{0c}$, together through a parent distribution, expressing the prior belief that there is a single dominant strain of \cov\ globally. 

Our primary goal in this study is to examine worldwide data to investigate whether climate and the BCG vaccination prevalence are important drivers of \cov\ spread. Because of the danger of confounding variables we used the base growth rate, $G_{0c}$ for each country in a machine learning feature extraction algorithm to pick the most important additional variables to include in our computationally intensive hierarchical Bayesian analysis. This lead us to exclude  \pmt\ pollution. In addition we undertook a separate analysis including A+ blood type. The data for blood type came from unpublished online sources and is therefore kept separate since it is less trustworthy.  

The remaining four most promising environmental factors were Temperature (T, $^\circ$C), Relative Humidity (RH, \%), BCG vaccination coverage (BCG, \%), Ultra-Violet Index (UVIndex); the latter included as a proxy for vitamin D production. To this set we added two testing-related variables to serve as diagnostics for potential contamination from testing issues: (1) the fraction of tests that return positive (Pos-Rate, \%) and (2) the number of tests per 1,000 population (Tests/1k), yielding our final set of global parameters, $\mathbf{\Theta}\equiv$ (T, RH, BCG, UV, Pos-Rate, Tests/1k). Due to concerns over data integrity, we study the impact of A+ blood type separately in section (\ref{sec:blood}). 

\begin{figure}[!ht]
    \centering
    \includegraphics[width=0.5\textwidth]{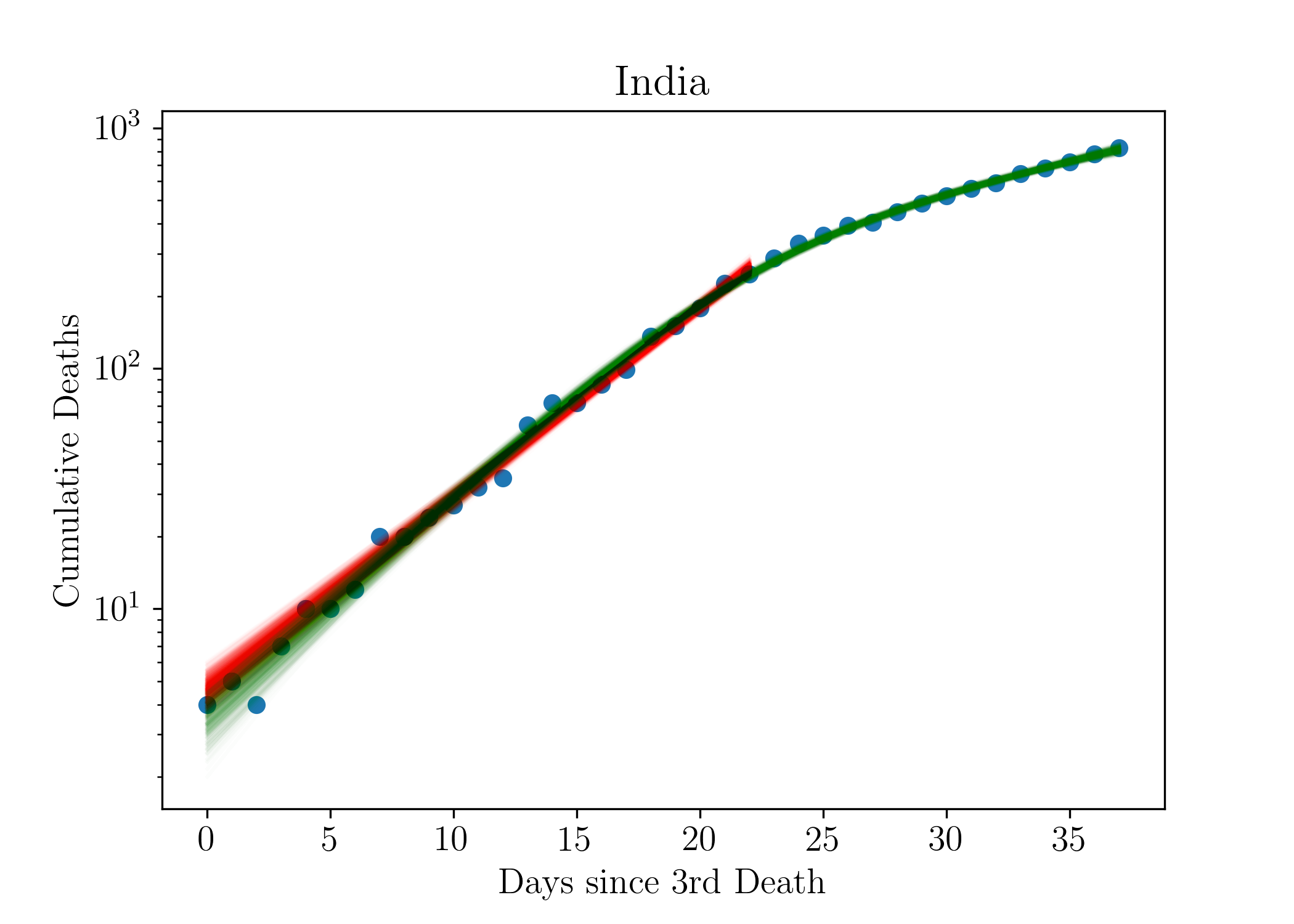}
    \caption{Example fit to the log of deaths vs time for India, showing how we select the initial linear part of the logarithmic data for inclusion in our analysis. Data are shown as blue points. Data covered by red fits are those used in our main analysis, in this case at $t_{0c} \leq 22$ days after reaching three deaths. Green samples are drawn from the 5d sigmoid model which covers all the data and models the changes in the growth rate due to social interventions or nonlinear effects. This is one of the 40 countries/provinces in our analysis; the full set is shown in Fig. (\ref{fig:example_fits}). }
    \label{fig:india}
\end{figure}

We then explicitly fit for $\mathbf{\Theta}$ using our sample of \numcountries\ regions in 37 countries for which there is data for all of the parameters in $\mathbf{\Theta}$.  We cannot use all the death data for countries since a constant growth rate model fails quickly for most countries due to Non-Pharmaceutical Interventions (NPIs) or nonlinear effects. To address this we extract the initial pure exponential part of the data on a country-by-country basis, as illustrated in Fig. (\ref{fig:india}) and discussed in detail in section (\ref{sec:additional}), which is the only data we use in our main analysis. 

The early phase death data for each country $c$, $D_c(t)$, are then fit to the model:
\begin{equation}\label{eq:deaths_with_factors}
    D_c(t) = D_{0c} \prod_{\tau=1}^t  \bigg(1 +  \beta \{G_{0c} + \mathbf{\Theta} \cdot \mathbf{X_c}\} \bigg)
\end{equation}
where $\beta = 10^{-2}$ converts our growths to percentages and the $\mathbf{X_c}$ are the country-specific data corresponding to $\mathbf{\Theta}$. 

We use a hierarchical Bayesian analysis in which each country's growth rate is drawn from a parent distribution which allows each country's base growth rate, $G_{0c}$, to be intrinsically different rather than forcing any differences to be due only to the parameters in the factors encoded by $\mathbf{\Theta}$.  

We use Monte Carlo Markov Chain (MCMC) methods to  simultaneously fit the base growth rates, $G_{0c}$, for all countries, our parameters of interest, $\mathbf{\Theta}$, and the parent distribution hyperpriors, implying a large Bayesian hierarchical model with 90 parameters in total. After marginalising over all country growth rates and parent distribution parameters, we are left with the marginal distributions on the $\mathbf{\Theta}$, our parameters of interest, providing our main results.  

We assess the importance of each of the $\mathbf{\Theta}$ parameters both through standard model selection metrics such as the Bayesian evidence and Bayesian Information Criterion (see Table \ref{tab:model_comp}) and by computing the statistical significance with which the parameters deviate from zero in the marginalised posterior chains. We now discuss these results. 

\section{Results and Discussion}

\begin{figure}[!ht]
    \includegraphics[width=0.55\textwidth]{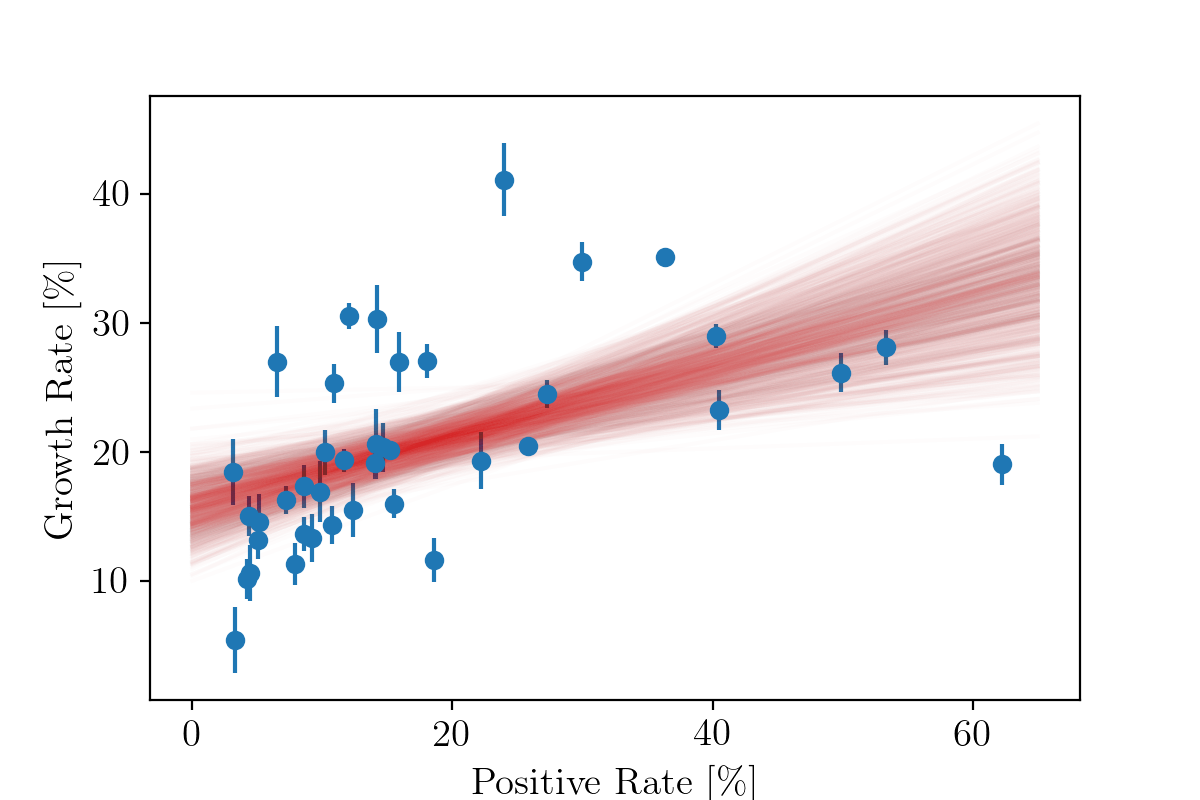}
    \caption{The growth rates with errorbars for the 40 regions taken from our No Factor model plotted against the positive rate data, are shown in blue. The red lines are posterior samples from the univariate positive rate model. The intercept was taken to be the parent level mean of the growth rate and the slope is the coefficient for positive rate of the same model; showing why this variable is detected at $3\sigma$.}
    \label{fig:pos_rate}
\end{figure}

Figure (\ref{fig:marginals_all_variables}) shows the main results of our paper: only the parameter associated to the positive rate or fraction of tests (Pos-Rate) is non-zero at more than $3\sigma$. This result is stable to significant changes in our priors and hyperpriors and to including or removing the other parameters in $\mathbf{\Theta}$ in the analysis. In addition, the positive rate is selected as the most important variable by all model-selection metrics (see Table \ref{tab:model_comp}) and by our machine learning analysis, section (\ref{sec:ml}).  A plot of regional death growth rates versus positive rate is shown in Fig. (\ref{fig:pos_rate}) showing the clear correlation.   

The simplest explanation of this result is that regions that experience the most rapid spread of the disease, i.e. those with the largest $R_0$, were also the regions that were least able to keep up with testing demands and hence where the rate of Polymerase Chain Reaction (PCR) tests used for \cov\ returning positive were the highest on average. This is then not a cause, but rather a result of, high growth rate. We find that Pos-Rate correlates positively with Tests per 1000 population (see Fig. \ref{fig:corner_plot_incl_testing}): high death growth rate and growing positive rate likely spurred increased testing. Neither of these observations help us identify the cause of increased death growth rates, however. 

Temperature is the only other variable which is non-zero at more than $2\sigma$ in our multivariate fits (see section \ref{sec:multiuniv} for more discussion) and relative humidity the only other parameter non-zero at more than $1.5\sigma$, providing some weak evidence for climatic impact on the spread of the disease. Increasing temperature and humidity tends to decrease the spread of \cov, in agreement with previous studies. 

In our model-selection metrics, where we compare fits with one variable at a time, relative humidity is preferred over temperature by the Akaike Information Criteria (AIC) and Bayesian Information Criteria (BIC) and is the only variable other than Pos-Rate, that has a BIC value more than 2 units lower than the No Factor model (which has $\mathbf{\Theta} = 0$)\footnote{$\Delta$ BIC$ = 2-6$ is the standard demarcation of positive evidence \cite{bic}.}.   The remaining parameters have poor or mixed results relative to the No Factor model, and all perform very poorly relative to the model with just the positive rate.

\begin{figure}[!ht]
    \centerline{
    \includegraphics[width=1.2\linewidth]{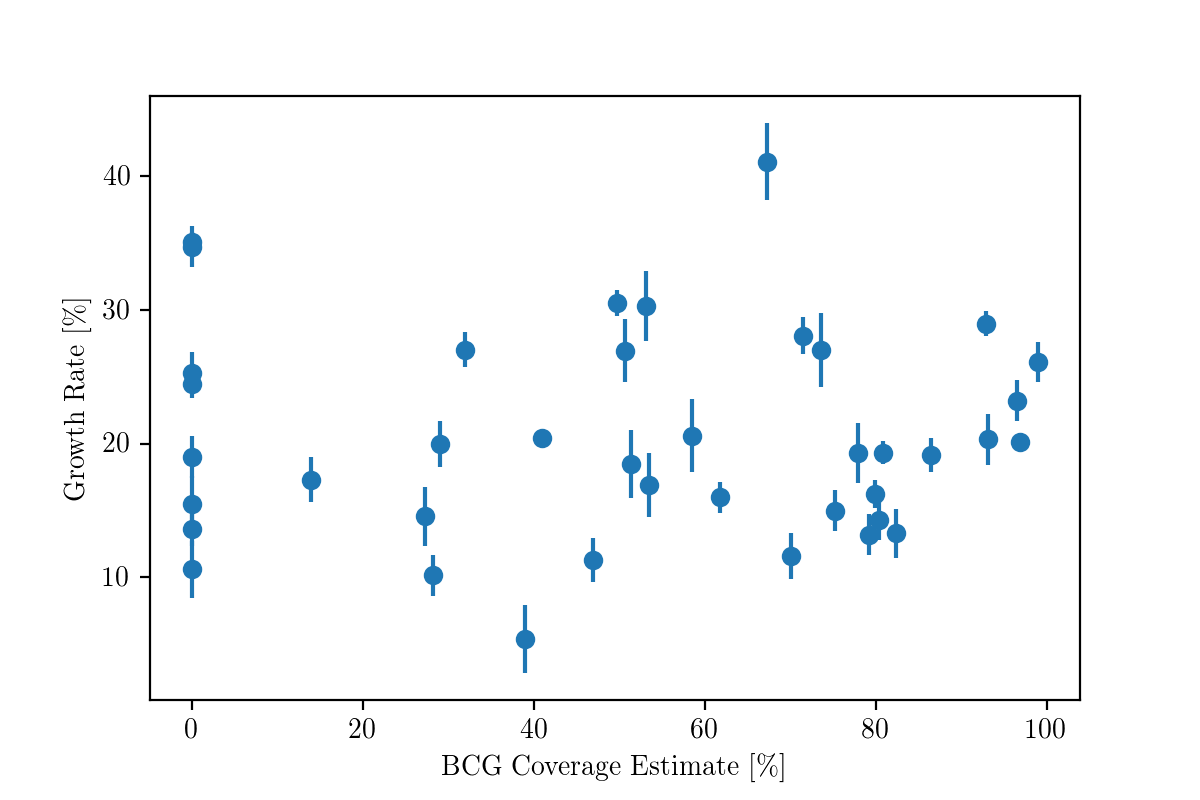}}
    \caption{BCG population coverage estimate against our estimated base growth rates. We see that there is no clear trend with increasing BCG coverage. The growth rates with error bars are taken from our No Factors model fit.}
    \label{bcg_growth}
\end{figure}

In particular, our regression and machine learning analyses provide no evidence in support of BCG fraction. This is not surprising if we look at figure (\ref{bcg_growth}) which plots regional death growth rates versus BCG coverage: there is no discernible correlation. Our results are therefore in agreement with \cite{bcg_neg,bcg_neg2}. Further we do not find any correlation between UV index and death growth rate. This is pertinent since UV index is relevant to natural production of vitamin D \cite{uv_vitd} and vitamin D has been suggested as a protective factor against the spread of the disease \cite{vitd,vitd2,vitd3}, 

Now we briefly discuss results for the other testing variable, Tests / 1k. We find that this is not a significant correlate of growth rate. This is perhaps not unexpected. If testing efficiency in each country is approximately constant, i.e. the rate of true infections detected changes only slowly with time, then this has little impact on the growth rate of deaths. It is only if the testing efficiency varies rapidly with time that we would expect this to be significant. Our results suggest that this was not the case, at least within the initial phase of the epidemic in each countries. Neither the Bayesian, nor the machine learning analyses found the number of tests per 1000 population to be significant.  We present technical details of the data, algorithms and analysis in section (\ref{sec:additional}).

Finally, in our separate analysis of blood types discussed in sections \ref{sec:blood} and \ref{sec:ml} we analyse the potential correlation of blood type with growth rate. We find marginal evidence for A+ blood type being relevant, subject to the caveats discussed in section (\ref{sec:blood}). 

How should our results be taken in the context of the many claims of climate, blood, BCG etc... being significant factors in \cov\ spread? First, many studies were based on confirmed \cov\ test cases which, as we discussed earlier, are affected by differences in testing capability and protocols between countries. 


Secondly, many of the studies present regressions that do not allow for unmodelled confounding sources of variation in the growth. Hence, if a country shows high growth the algorithm will try to force one of the potential factors under study to explain it. Instead the hierarchical Bayesian framework allows the base growth rate of each country to be different, and hence potential factors will only be given credit for the difference in the growth rate if they provide a genuinely better fit. 

Further, many studies do not model the intrinsic uncertainties associated with the data as we have done. We too find that the best-fits are non-zero (as can be seen by looking at the peaks of the posteriors in Figure (\ref{fig:marginals_all_variables}) or at the last rows of Table (\ref{tab:all_results}). Hence our results are not in disagreement with regression results, the issue is about the statistical significance of such claims. 

Finally, although we do not detect environmental factors at more than $3\sigma$, it is interesting to examine how big the environmental factors would be if our best-fit parameter values describe reality: a $15^{\circ}C$ increase (decrease) in temperature implies a $5.25\%$ decrease (increase) in the base daily growth rate while a increase (decrease) of 20\% in relative humidity would mean a $2.24\%$ decrease (increase) in base daily growth rate. The decreased spread at higher temperature and humidity agrees with previous work \cite{weather2}. 

This may not seem significant, but for a city such as Johannesburg, where both temperature and humidity drop significantly in winter, the combined effect could add more than $5\%$ to the base daily growth rate. For a daily growth rate of $5\%$, which was approximately the value in May 2020, this would halve the doubling time of the disease, a significant impact.

\section{Conclusions}

Contrary to previous claims our analysis of growth rates for deaths from countries worldwide are consistent with no effect from climate, pollution or BCG vaccination. The only significant correlation detected is with the positive rate of tests: a country that intrinsically had a high $R_0$ (due to high population density etc...), would naturally tend to be more overwhelmed and hence run low on testing kits earlier, leading to increased fraction of positive tests. We did find some weak suggestive evidence, at $< 3\sigma$, that temperature and relative humidity correlate with death growth rates. 

A separate analysis of blood type data shows that A+ type is the most important blood type, though the significance is marginal, both because the data quality is low and the statistical significance is weak.  Our combined statistical and machine learning analysis finds no evidence for \pmt\ pollution, other blood types or UV Index as drivers of \cov. 

More data could be obtained by dropping the requirement that all countries in the sample have data for all the potential factors, which could potentially allow for some of the effects to be detected at higher statistical significance but at the cost of making model comparison significantly more difficult. 


\vskip 2mm 
The data and  code for our Bayesian analysis is available at https://github.com/chrisfinlay/covid19/.

\subsection{Acknowledgements}

We thank Michael van Niekerk for extracting data from BCG ATLAS,  Niayesh Afshordi, Ewan Cameron,  Inger Fabris-Rotelli, Ben Holder and Nadeem Oozeer for discussions and comments.  

This research has been conducted using resources provided by the United Kingdom Science and Technology Facilities Council (UK STFC) through the Newton Fund and the South African Radio Astronomy Observatory.

\begin{figure*}[!ht]
    \vspace*{-1cm}
    \centerline{
    \includegraphics[width=1.2\linewidth]{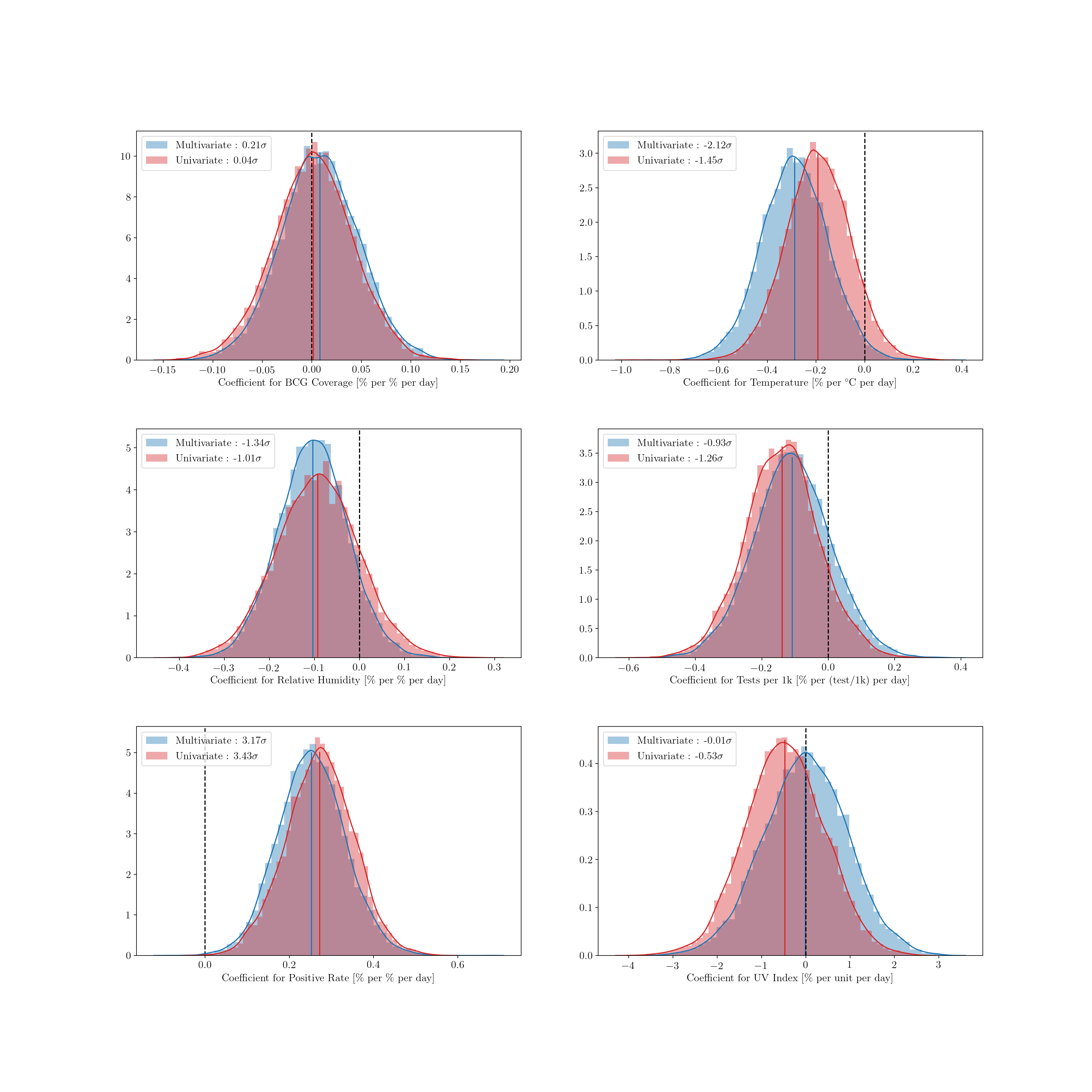}}
    \vskip -1.5cm
    \caption{Marginal distributions for the various factors we consider in the case where all factors are considered simultaneously (blue; multivariate) and individually (red; univariate). The results in both cases are consistent: only the coefficient associated with the positive rate of tests is non-zero at more than $3\sigma$, with temperature non-zero at $2\sigma$ in the multivariate case. Notice that the significance of both the temperature and relative humidity increase in the multivariate fits relative to the univariate fits where they are fit alone. Both BCG and UV Index are consistent with no effect. }
    \label{fig:marginals_all_variables}
\end{figure*}

\begin{table}
    \centering
    \begin{tabular}{lccc}
    \hline
    \hline
	Model & \bf{$\Delta$AIC} & \bf{$\Delta$BIC}  & \bf{$\Delta$DIC}  \\ 

	\hline
    \hline
    \\
	Positive Rate  &    0.0    &    0.0    &    0.0   \\ 
	Relative Humidity  &    5.5    &    5.5    &   31.8   \\ 
	No Factors  &   15.3    &   11.6    &   22.9   \\ 
	Temperature  &   13.7    &   13.7    &   24.0   \\ 
	BCG Vaccine  &   14.7    &   14.7    &   13.6   \\ 
	UV Index  &   14.7    &   14.7    &   30.5   \\ 
	Tests / 1k  &   15.2    &   15.2    &   28.6   \\ 
	All Incl.Tests  &  114.5    &  132.9    &  113.2   \\ 
	All Excl. Tests  &  125.9    &  137.0    &  127.4   \\ 



\\
	\hline
	\hline
    \end{tabular}
\caption{Model selection rankings relative to the best model (first row) for four metrics estimated from the MCMC chains: Akaike Information Criteria (AIC), Bayesian Information Criteria (BIC) and Deviance Information Criteria (DIC). Here we consider each factor in $\mathbf{\Theta}$ separately (univariate), as well as the model with no $\mathbf{\Theta}$ factors, the model with all factors and a model with all factors excluding the two testing factors (Positive Rate and Tests / 1k). The Positive Rate is unanimously selected as the most important feature, followed by the Relative Humidity. Temperature does not perform well here but this is not surprising: the statistical significance of the temperature increased by $0.84\sigma$ in the joint multivariate fit relative to the univariate fit alone.A similar increase in significance is visible for the Relative Humidity; see Fig. (\ref{fig:marginals_all_variables}). Note that the model including the two testing parameters (All Incl. Tests) outperforms the model excluding the testing parameters, reinforcing the fact that testing is important. A+ blood type results are presented separately in section (\ref{sec:blood}).}
    \label{tab:model_comp}

\end{table}

\newpage

\section{Additional Material}\label{sec:additional}

\subsection{Data}

\subsubsection{Data on deaths}\label{sec:t0}
We chose to look at the growth rate in deaths as they are less likely than confirmed cases to be affected by the widely varying testing protocols between countries.  The cumulative death data for each country comes from the Centre for Systems Science and Engineering (CSSE) at Johns Hopkins University (JHU) \cite{covid_data}.

We cannot, however, take all data on deaths for all times since it is clear that a simple exponential model fails quickly: as soon as interventions occur the simple exponential model fails and we will obtain contaminated estimates of the growth rate due to the flattening of the curve which will skew our analysis; see e.g. Fig. (\ref{fig:india}). As a result we want to know which data should be included in our analysis for each country. This reduces to knowing the time, $t_{0c}$, at which the growth of deaths deviates from a simple exponential. We then only use data at $ t < t_{0c}$ for country $c$. 

Doing this cut by hand introduces the possibility of human bias.  Instead we compute $t_{0c}$ for each country by fitting a non-linear sigmoid transition function to the growth rate of the death data of each country over time:
\begin{equation}
    \sigma_c(t;t_{0c}, \alpha_c) = \frac{1}{1+\exp \big[-\alpha_c (t-t_{0c}) \big]}
\end{equation}
This function allows the growth rate to smoothly transition from an initial value (which we are interested in) to a final value, reflecting the impact of social interventions or nonlinearity in the system. 

In this first phase of the analysis each country, $c$, therefore has 5 parameters ($D_{0c}, G_{0c}, \delta G_{c}, \alpha_c, t_{0c}$) to describe the trajectory of its deaths, where $\delta G_{c}$ and $\alpha_c$ represent the change in growth rate and suddenness of the transition with time from $G_{0c}$ to $G_{0c} + \delta G_{c}$ for each country, $c$ (the index, $c$, has been left out in Eq. (\ref{sigmoid_fit}) below):  
\begin{equation}
    D(t; D_0, G_0, \delta G, \alpha, t_0) = D_0 \prod_{\tau=1}^t \big(G_0 + \delta G \cdot \sigma(\tau; t_0, \alpha) \big)
    \label{sigmoid_fit}
\end{equation}
To maximise the probability that this is a good fit to the data the non-linear model in Eq. (\ref{sigmoid_fit}) we only consider data up to the time when a country passes 1000 deaths. All five parameters are treated hierarchically with their own parent distribution where the means are Normally distributed and the standard deviation is HalfNormal distributed. 

We only use data from $t < t_{0c}$ in our main analysis (see section \ref{model}), where $t_{0c}$ is given by the marginalised mean from the chains. Typical values for $t_{0c}$ were around 15 days. We excluded countries where $t_{0c} < 10$ because of concerns about quality of the underlying model fit in such cases, i.e. a simple exponential model was not a good fit even at early times which could lead to spurious growth rates.  

\begin{figure*}[ht]
    \centering
    \includegraphics[width=1.0\textwidth]{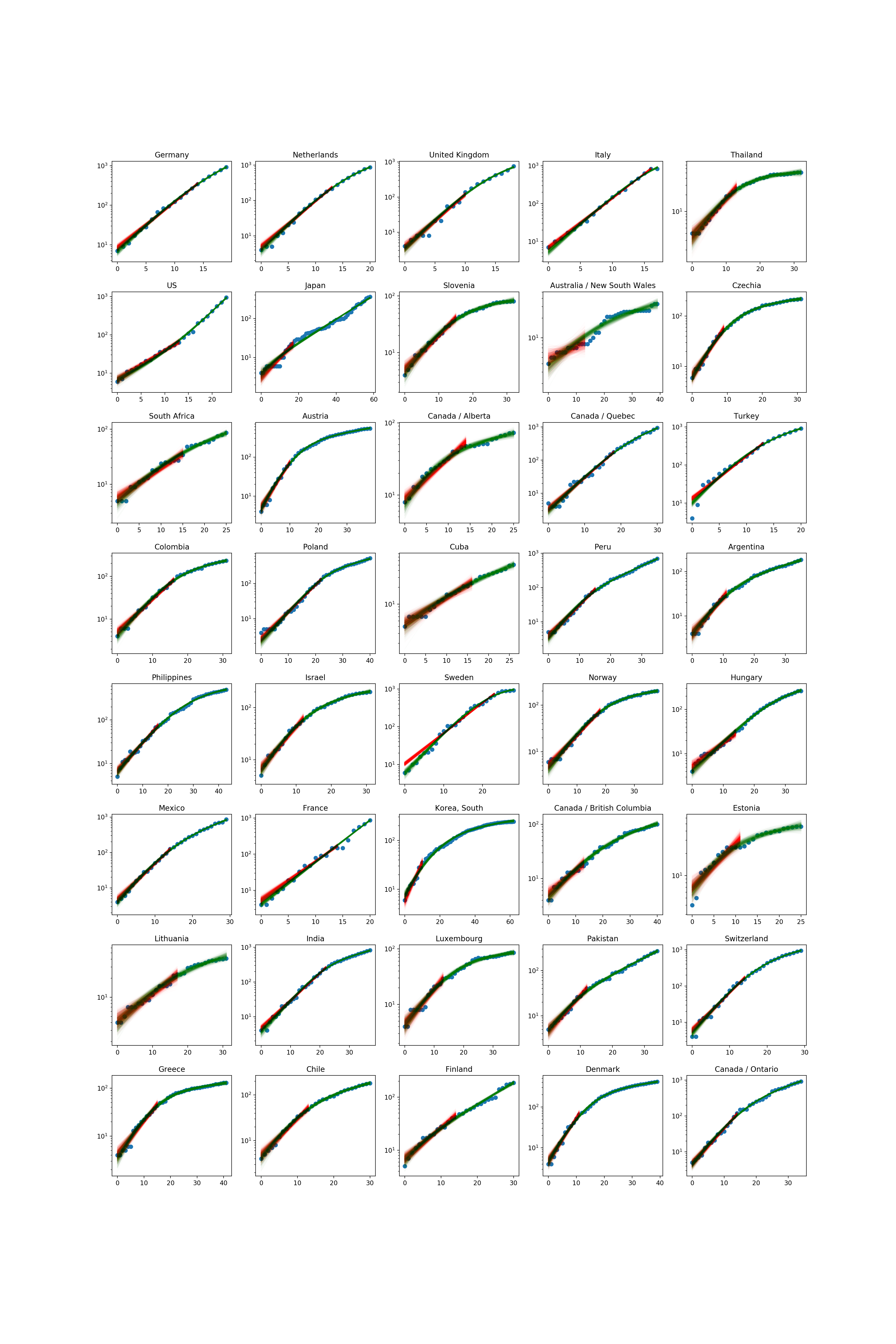}
     \vskip-3.3cm
    \caption{Example fits to all 40 regions from 37 countries in our data sample. Posterior samples from the simple exponential model is shown in red, plotted up to the cutoff date, showing the data we use in our analysis. The full model with time varying growth rate is shown in green while data points are shown in blue. Only the USA had an exponent larger after $t_{0c}$ than before. }
    \label{fig:example_fits}
\end{figure*}

The resulting fits to country death data for both the full 5-d model (green) and just the region $t < t_{0c}$ are given in Figure (\ref{fig:example_fits}), showing that the technique performs well in isolating the initial exponential growth from a first-principles approach. 

Once we have $t_{0c}$ for each country we further cut our data by using the following rules: 
\begin{itemize}
    \item We only use countries for which there were a minimum  of 20 deaths by 25 April 2020.
    \item We choose day zero as the first day a country passes 3 deaths. This leads to typical starting numbers of deaths of around 5.
    \item Data for our variables of interest (Climate, BCG, etc...) were not available for all countries. To ensure the same amount of data for all models only countries with data for all our parameters in $\mathbf{\Theta}$ were included.
  \end{itemize}
  
We did not model the potential correlations and interactions of our parameters with these data cuts. This could potentially alter our conclusions: if a parameter were extremely important in determining growth rates, then countries with very small numbers of deaths would systematically not make it past our data cuts and hence the signal from that data would be lost. However, it is unclear how to model this censorship rigorously and it is left to future work.
  
After these data cuts we were left with 40 provinces in 37 countries with 613 data points in total, shown in Figure (\ref{fig:example_fits}). The only country with more than one province was Canada which include Alberta, Ontario, Quebec and British Columbia (BC).  

\subsubsection{Climate Data}

For each region and country the climate data was gathered from the Dark Sky API, \cite{climate_data}, where the locations sampled are taken from the latitudes and longitudes given for each country or region in the JHU \cov\ data. For each country, $c$, the mean temperature, mean relative humidity and mean UV Index are calculated as an average over a N day window starting 28 days prior to day 0 for each region, where N is the number of days of deaths data for that country. The latter is chosen as an estimate of the average time from infection to death \footnote{https://www.thelancet.com/journals/laninf/article/PIIS1473-3099(20)30243-7/fulltext}. Since mean climate variables change relatively slowly the exact delay is not important. See Fig (\ref{fig:example_fits}) for the distribution of $t_{0c}$ times. We used UV Index as a proxy for natural vitamin D production. Data for all regions and countries is shown in Table (\ref{allresults}).

\subsubsection{BCG Vaccine Coverage}

Bacillus Calmette–Guérin (BCG) vaccination policies have been in place in many countries across the world starting from widely differing start dates. We would like to estimate the percentage of the population in each country that has received a BCG vaccine in their lifetime. For this purpose we need the age demographics of each country as well as the dates when BCG vaccinations became/stopped being mandatory.

To estimate the percentage of the population vaccinated by BCG we need to draw on three sources of data. These are the BCG Atlas, World Health Organisation (WHO) BCG vaccination rates amongst 1-year olds and the age demographics for each country from the United Nations (UN). Firstly we looked to the BCG Atlas \cite{bcg_data1}. This is a heterogeneous dataset which implies that not all information was available for each country. We collected the following fields from the BCG Atlas:
\begin{enumerate}
    \item Current BCG vaccination?
    \item Which year was vaccination introduced?
    \item Year BCG stopped?
    \item Year of changes to BCG schedule
    \item Details of changes
    \item BCG coverage (\%)
\end{enumerate}
We did not use the last field directly due to missing information on this field, specifically if it was for a small age range of the population or its entirety. For some countries data was missing from fields 1-3. In these cases, where appropriate and possible, the missing data was obtained from fields 4 and 5. 

Age demographics for each country were used to compute the expected fraction of the population who have had the BCG vaccination. In 15\% of cases the BCG Atlas did not have data and we instead used WHO data \cite{bcg_data2} to perform the estimate. The derived BCG fractions and the origin of the data, are shown in Table (\ref{bcg_fraction}) while a plot of the BCG fractions versus growth rates are shown in Fig. (\ref{bcg_growth}). 

There is one caveat here: our BCG coverage is the estimate of the percentage of the population of each region and country that has had the BCG vaccination. This is arguably not the optimal quantity to use in our analyses however; it might be better to use the fraction of vaccinated population weighted by the probability of infection as a function of age. However, the latter is currently unknown and hard to compute even in the best of situations: how can we know how many people were exposed but never got infected? The large scatter in Figure (\ref{bcg_growth}) suggests that this is unlikely to make a significant difference to our conclusions that BCG vaccine is not important  in the spread of  \cov.

\begin{table}
    \centering 
    \begin{tabular}{llc}
        \hline
        \hline
	Country & BCG Coverage  & Data Source  \\ 
	\hline
	\hline
    %
    %
    \\
    Argentina & 58.5 \% & WHO \\     
    Australia / NSW & 38.9 \% & BCG ATLAS \\     Austria & 53.1 \% & BCG ATLAS \\     
    Canada / Alberta & 0.0 \% & BCG ATLAS \\    
    Canada / BC & 0.0 \% & BCG ATLAS \\     
    Canada / Ontario & 0.0 \% & BCG ATLAS \\    
    Canada / Quebec & 0.0 \% & BCG ATLAS \\     
    Chile & 93.1 \% & BCG ATLAS \\     
    Colombia & 86.4 \% & BCG ATLAS \\     
    Cuba & 46.9 \% & WHO \\     
    Czechia & 73.5 \% & BCG ATLAS \\     
    Denmark & 50.7 \% & BCG ATLAS \\     
    Estonia & 27.2 \% & WHO \\     
    Finland & 80.4 \% & BCG ATLAS \\     
    France & 71.4 \% & BCG ATLAS \\    
    Germany & 49.7 \% & BCG ATLAS \\
    Greece & 13.9 \% & WHO \\     
    Hungary & 82.3 \% & BCG ATLAS \\     
    India & 96.9 \% & BCG ATLAS \\     
    Israel & 29.0 \% & BCG ATLAS \\     
    Italy & 0.0 \% & BCG ATLAS \\     
    Japan & 70.0 \% & BCG ATLAS \\     
    Korea, South & 51.3 \% & BCG ATLAS \\     
    Lithuania & 28.2 \% & WHO \\     
    Luxembourg & 0.0 \% & BCG ATLAS \\     
    Mexico & 98.9 \% & BCG ATLAS \\     
    Netherlands & 0.0 \% & BCG ATLAS \\     
    Norway & 79.8 \% & BCG ATLAS \\     
    Pakistan & 77.8 \% & BCG ATLAS \\     
    Peru & 96.5 \% & BCG ATLAS \\     
    Philippines & 61.8 \% & WHO \\     
    Poland & 80.8 \% & BCG ATLAS \\     
    Slovenia & 75.2 \% & BCG ATLAS \\     
    South Africa & 79.2 \% & BCG ATLAS \\     
    Sweden & 40.9 \% & BCG ATLAS \\     
    Switzerland & 31.9 \% & BCG ATLAS \\     
    Thailand & 53.4 \% & BCG ATLAS \\     
    Turkey & 92.8 \% & BCG ATLAS \\     
    US & 0.0 \% & BCG ATLAS \\     
    United Kingdom & 67.2 \% & BCG ATLAS \\ 
    \\
	\hline
	\hline
    \end{tabular}
    \caption{BCG vaccine coverage for populations for all our regions/countries estimating the fraction of the population who have received the vaccine. }
        \label{bcg_fraction}
\end{table}

\subsubsection{Additional data}
Data on prevalence of blood type for each country was taken from \cite{bloodtype1,bloodtype2} while \pmt\ pollution data came from \cite{pm25}. We discuss the blood type data and results in section (\ref{sec:blood}). 

\subsection{Model} \label{model}

As described earlier our basic regression model for the deaths in country $c$ at day $t$ is: 

\begin{equation}
 D_c(t) = D_{0c} \prod_{\tau=1}^t \bigg( 1+\frac{G_c}{100} \bigg)\label{eq:deaths_G}
\end{equation}
which is assumed valid for $T \leq t_{0c}$, as described before. $G_c$ is measured in percent and we model it's potential dependence on our variables of interest as:
\begin{equation}
    G_c(\mathbf{\Theta}) = G_{0c} + \mathbf{\Theta} \cdot \mathbf{X_c}
\label{eq:deaths_with_factors2}
\end{equation}
where $G_{0c}$ and $\mathbf{X_c}$ are the country-specific base growth rate and data for climate, BCG etc..., shown in Table (\ref{allresults}) and $\mathbf{\Theta}$ are the global parameters we are interested in. In general the $\mathbf{X_c}$ could be time-dependent. In this analysis, because of missing data and the fact that our data for each region typically spans a short period ($t_{0c}$ is less then 3 weeks as discussed in section \ref{sec:t0}), we use average values for $\mathbf{X_c}$.

\begin{figure}[!ht]
    \centerline{
    \includegraphics[width=1.0\linewidth]{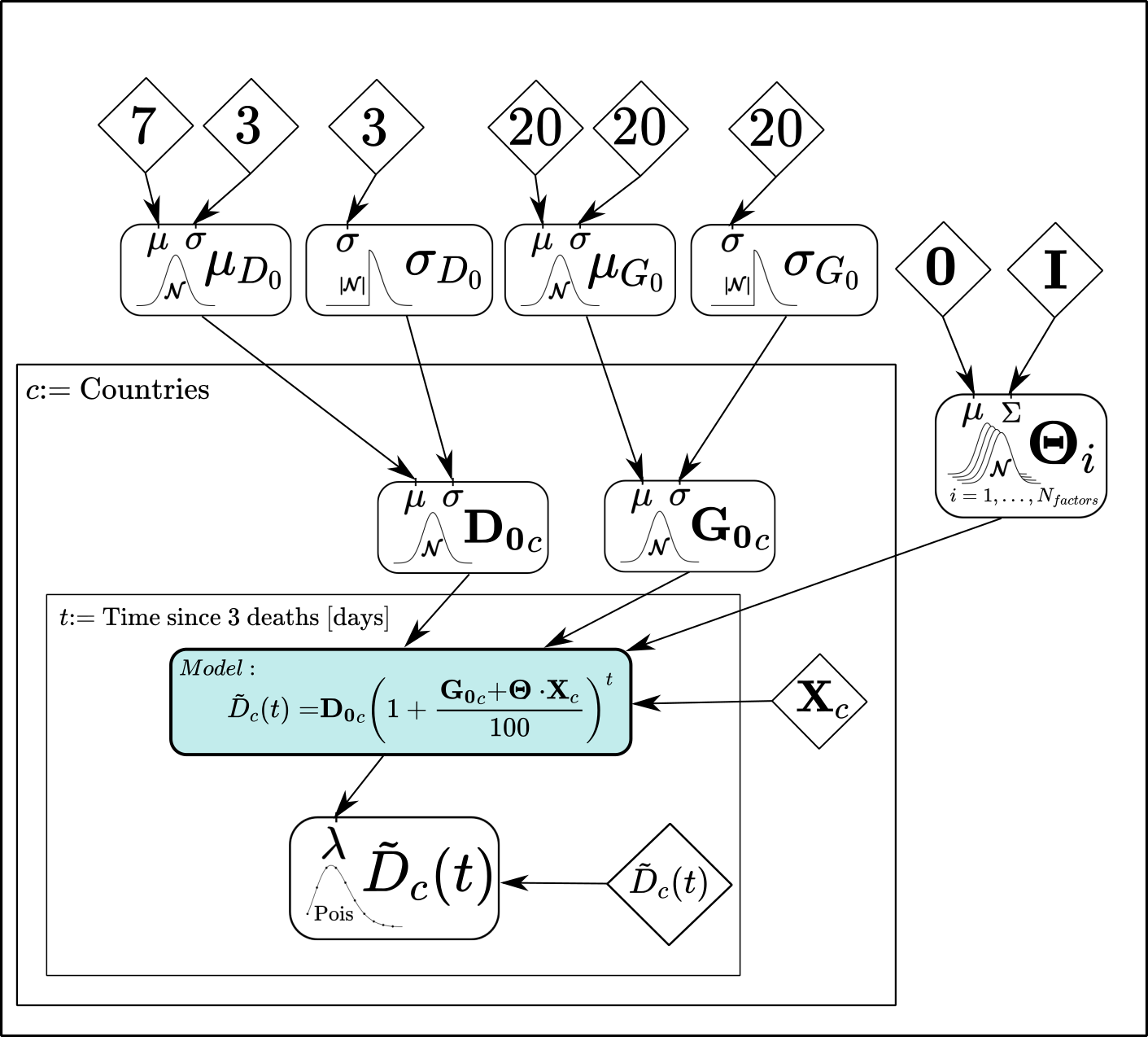}}
    \caption{Graphical model for our hierarchical Bayesian analysis showing the parameters, data and their connections. Rectangles with rounded corners represent parameters in the model with their priors shown as normal or half normal distributions with their respective parameters. Diamonds are fixed inputs/data. Squares represent repetition plates with repeated variables with index given in the top-left corner of the plate. $\tilde{D}(t)_c$ are the observed deaths for country $c$ on day $t$.}
    \label{fig:platemodel}
\end{figure}

Our goal is to determine if any of the parameters $\mathbf{\Theta}$ are rigorously required to be non-zero by the data. One limitation of Eq. (\ref{eq:deaths_with_factors2}) is that it is linear in the country data $\mathbf{X_c}$. We justify this by noting that the growth rates and the underlying data vary over narrow ranges, so that retaining only the linear terms in the Taylor series expansion of $G_c(\mathbf{\Theta})$ is a reasonable step. For the temperature variables we have verified that assuming instead a step change in the growth at around $10^{\circ} $C with two hierarchical growth parameters, one on each side of the step, did not lead to any increase in significance in the detection of temperature as an effect. 

\subsection{Posterior model and sampling} \label{sec:probsample}

We write a hierarchical Bayesian probabilistic model by assuming a Poisson likelihood, suitable for count data, with mean given by the deterministic forward model defined in Eq. (\ref{eq:deaths_with_factors2}), together with priors and hyperpriors for all our 90 parameters, as shown in the schematic graphical representation in Fig. (\ref{fig:platemodel}); see e.g. \cite{gelman}. We do not try to model missing deaths due to testing irregularities. As long as the fraction of missing deaths remains approximately constant in the early phase of the spread within the country that we consider this should have little impact on our results. 

Since our data covers 37 countries and 40 provinces worldwide and we know that the growth rate of the disease will depend both on properties of the disease (which are universal), and properties specific to each country (e.g. culture and population density) it is natural to model the data with a hierarchical Bayesian structure which allows growth rates to vary somewhat from country to country but to also be somewhat similar between countries. 


The priors and hyperpriors for each variable to be estimated is chosen to be: 
\begin{itemize}
    \item $\mu_{D_0} \sim \mathcal{N}(7, 3^2)$; the hyperprior for the mean on the prior for initial deaths, $D_{0c}$, for each country ($D_{0c} > 3$ as part of our data cuts).
    \item $\sigma_{D_0} \sim \text{HalfNormal}(3)$; the standard deviation on the prior for initial deaths for each country. 
    \item $\mu_{G_0} \sim \mathcal{N}(20, 20^2)$; the mean (in percent) on the prior for the growth rate of each country, $G_{0c}$.
    \item $\sigma_{G_0} \sim \text{HalfNormal}(20)$; the standard deviation on the prior for the growth rate of each country.
    \item $D_{0c} \sim \mathcal{N}(\mu_{D_0},\sigma^2_{D_0})$, the prior on the initial deaths parameter for each country, $c$. All countries together will be denoted by $\mathbf{D_0}$.
    \item $G_{0c} \sim \mathcal{N}(\mu_{G_0}, \sigma^2_{G_0})$; the prior on the growth rate parameter for each country, $c$. All countries together will be denoted by $\mathbf{G_0}$.
    \item $\mathbf{\Theta} \sim \mathcal{N}(\mathbf{0}, \mathbf{I})$; the prior on the vector of parameters of key interest.
\end{itemize}

The complete vector of all parameters is therefore:
\begin{equation}
    \big( \mu_{D_0}, \sigma_{D_0}, \mu_{G_0}, \sigma_{G_0}, \mathbf{D_0}, \mathbf{G_0}, \mathbf{\Theta} \big)^T
    \label{eq:allparams}
\end{equation}
The joint prior over our full set of parameters is assumed factorizable, i.e. a product of the prior distributions listed above.

We assume our data to be statistically independent both between countries and from day-to-day. Our model simultaneously fits the death data from all countries. We use NumPyro\cite{numpyro} for Monte Carlo Markov Chain (MCMC) probabilistic sampling from our posterior using the No U-Turn Sampling (NUTS) algorithm \cite{NUTS}. NumPyro has diagnostic tools built in and allows for easy running on accelerators, such as GPUs, which were key in being able to iterate quickly through our many models including all the multivariate and univariate combinations, as well as looking at the effect of priors.  

The MCMC simulations were generally run as 4 independent chains with each chain starting from a random position sampled from the prior for our parameters and run for 2000 steps in order for length scales (and other sampling options) to be determined automatically by NumPyro. After this initial ``burn-in" each chain was typically run for 5000 samples. Chains were long enough to ensure that the Gelman-Rubin convergence test was always less than 1.01. 

Each chain was then thinned by a factor of two in order to further improve independence of samples.
This leads to the final 10k samples collected for each model and led to good, converged, trace plots and posteriors; see Fig. (\ref{fig:marginals_all_variables}). We used ChainConsumer\footnote{https://samreay.github.io/ChainConsumer/} for the analysis of the chains, the model selection metrics in Table (\ref{tab:model_comp}) and some of the plots \cite{chainconsumer}.

\subsection{Additional Bayesian Results}

\subsubsection{A+ Blood Type Analysis}\label{sec:blood}

The data for blood type prevalence comes from online heterogeneous collections of published and unpublished data sources covering a wide range of publication dates. As a result, data integrity is an issue and we have chosen to separate the blood analysis from our main Bayesian analysis. 

A random forest analysis finds only A+ blood type as potentially relevant amongst the different ABO blood types (see section \ref{sec:ml}). Here we present the results of both univariate and multivariate A+ blood type Bayesian analyses using the formalism described in section (\ref{model}). 

\begin{figure}[!ht]
    \includegraphics[width=0.55\textwidth]{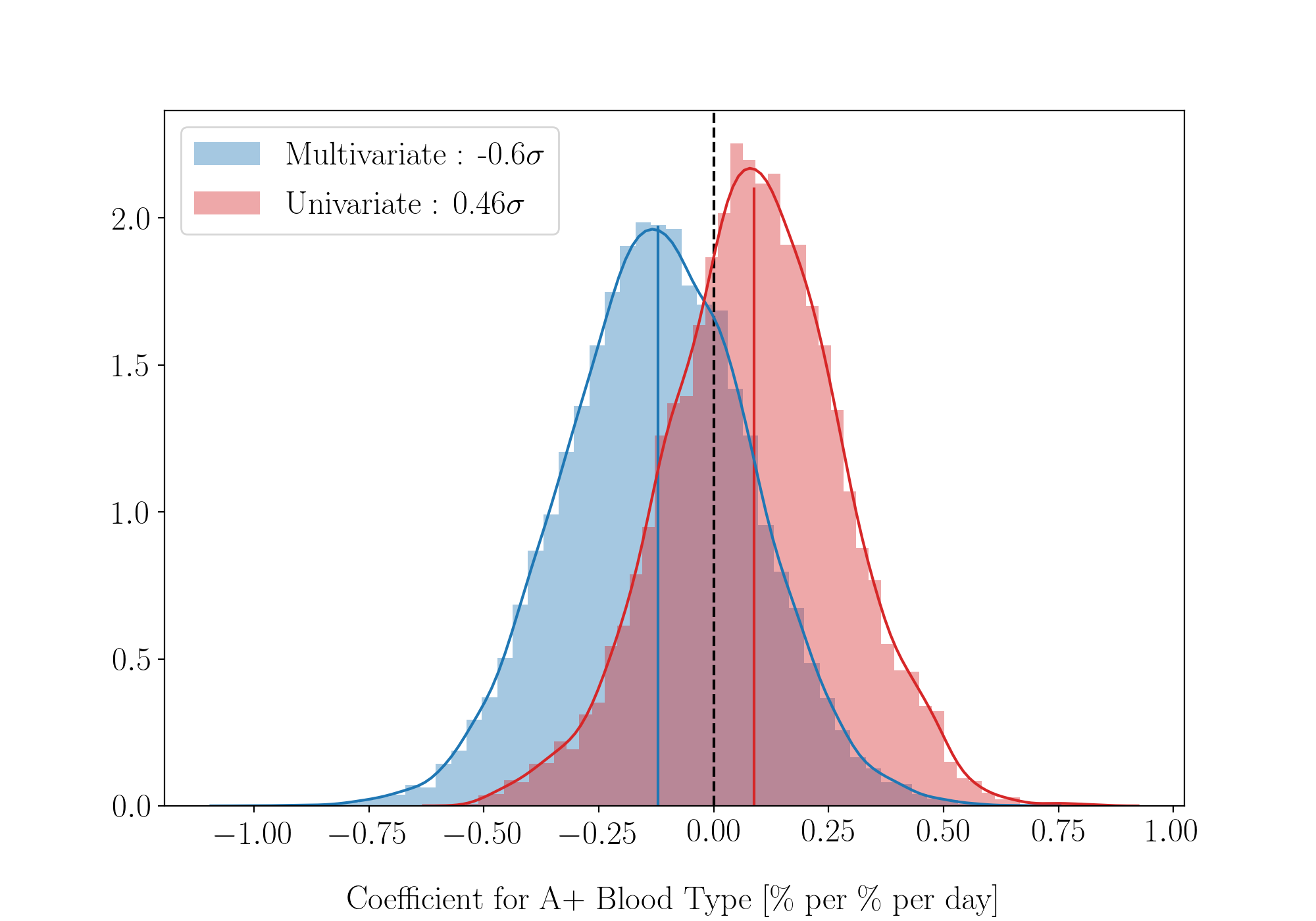}
    \caption{The univariate and multivariate results for the coefficient of A+ blood type are consistent with zero, i.e. no effect.}
    \label{fig:aplusblood}
\end{figure}

Fig. (\ref{fig:aplusblood}) shows the result of the Bayesian analysis for the A+ blood type coefficient: it is consistent with zero in both the univariate and multivariate cases. If we look at the AIC, BIC and DIC for A+ blood type we find values of: $(10.7,  10.7,   23.4)$ compared to the no-factor models results of $(15.3, 11.6, 22.9)$. The A+ AIC and BIC are better than the no-factor model, but the DIC value is worse. In addition, the random forest analysis found A+ blood type to be more important than both BCG and and Tests/1k. 

As a result we conclude that there is somewhat conflicting evidence regarding the potential effect of A+ blood type, but that none of the evidence is strong.  

\subsubsection{Effect of $\mathbf{\Theta}$ on Base Parameters}

In the results section we focused on the best-fit values of $\mathbf{\Theta}$. The reverse question is interesting too: what is the impact of including the $\mathbf{\Theta}$ parameters on the hyperprior parameters, mean, $\mu_{G_0}$, and standard deviation, $\sigma_{G_0}$, of the base growth rates, $G_{0c}$, for the 40 regions?
If the $\mathbf{\Theta}$ do explain some of the variation we would expect the standard deviation on the base growth rate (parent distribution) to get smaller, which is captured by the hyperparameter $\sigma_{G_0}$. The mean of the base growth rates (parent distribution), given by $\mu_{G_0}$, are nothing more than an offset dictating the value when all factors are zero i.e. $\mathbf{\Theta}=\mathbf{0}$.

This is what we see: in the No-Factor model $\sigma_{G_0} = 7.9\%$ which changes to $6.3\%$ when we allow all the $\mathbf{\Theta}$ to vary, a significant shrinkage.  On the other hand, if we include all parameters except for the two testing parameters,  $\sigma_{G_0}$ returns to $7.9\%$, while  $\sigma_{G_0} = 6.7 \% (7.7\%)$ when we add the positive rate (Tests/1k) parameters  alone; showing that it is primarily the positive rate parameter that drives the shrinkage in the uncertainty in the parent distribution on the base growth rates. 

\subsubsection{Comparison of univariate and multivariate fits}
\label{sec:multiuniv}
To assess the stability of our results we fit each of our key parameters in $\mathbf{\Theta}$ both alone (univariate) and simultaneously with all the other parameters (multivariate). As shown in Table (\ref{tab:uni_multi}) and Fig. (\ref{fig:marginals_all_variables}), the results are quite stable. Positive rate and temperature are the still the most significantly parameters in both cases: the positive rate is non-zero at more than $3\sigma$ in both cases while the significance for temperature increases from $1.45 \sigma$ to $2.19 \sigma$ when going from the univariate to multivariate fit.  Relative Humidity also increases in significance in the multivariate case. 

\begin{figure*}[ht]
    \centering
    \includegraphics[width=0.85\textwidth]{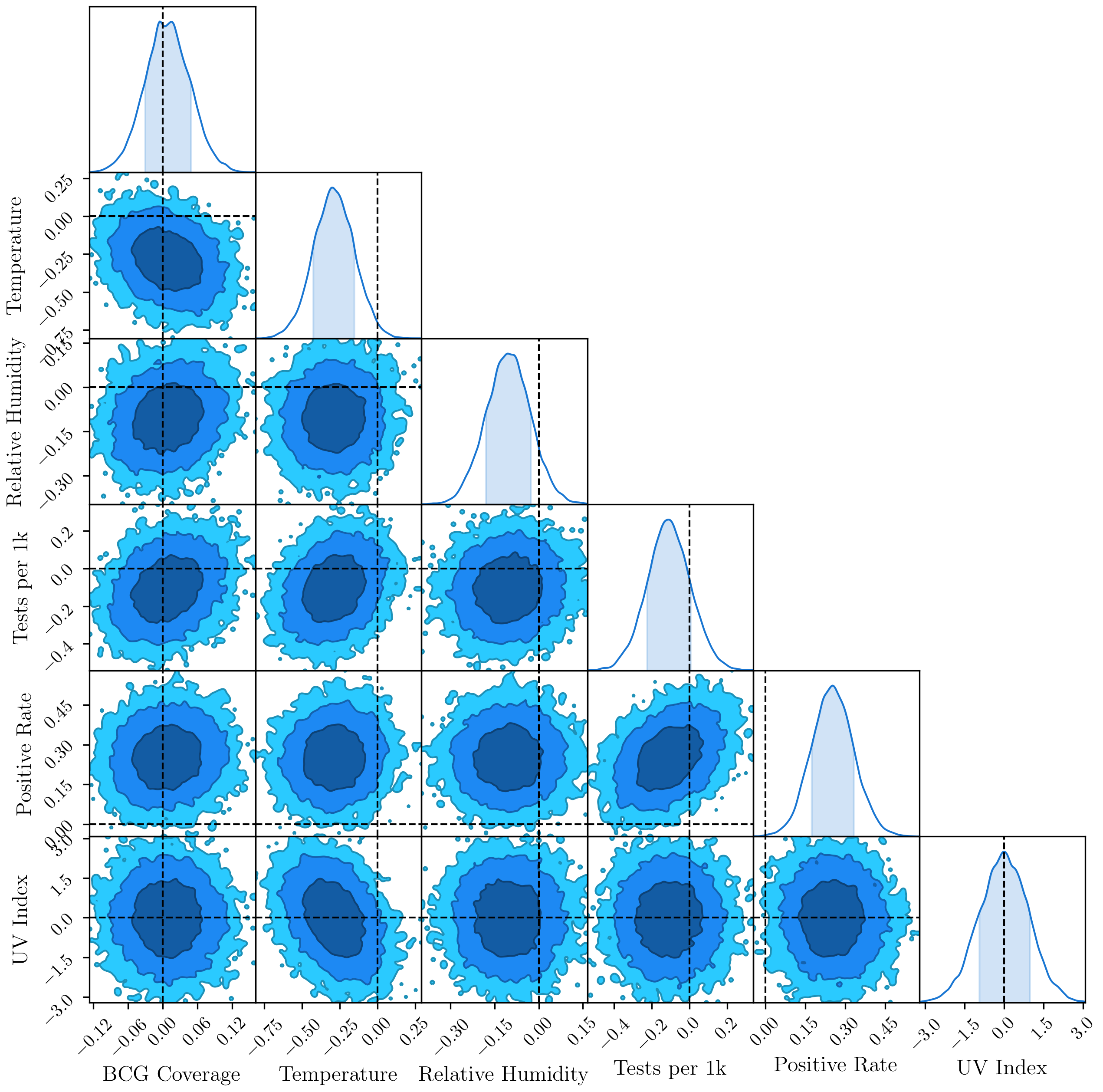}
    \caption{1, 2 and 3-$\sigma$ contours from the marginalised posterior samples of the parameters $\mathbf{\Theta}$ showing that most parameters are weakly correlated aside from one positive (Pos-Rate \& Tests/1k) and one negative (Temp \& UVIndex) correlation. Zero values for the parameters are shown by dashed lines to help assess statistical significance.}
    \label{fig:corner_plot_incl_testing}
\end{figure*}

To access correlations between parameters we show in Fig. (\ref{fig:corner_plot_incl_testing}) the one and two-dimensional marginalised posterior plots. Correlations are typically small, though there are is some small positive correlation between Tests per 1k and Positive Rate, and a small negative correlation between Temperature and UV Index.

Our full results for all the hyperpriors, the base country growth rates, $G_{0c}$ and the $\mathbf{\Theta}$ parameters for each of the different univariate and multivariate models are shown in Table (\ref{tab:all_results}). 

\begin{table}
    \centering
 \begin{tabular}{lrr}
\hline
\hline
Parameter & Multivariate   & Univariate  \\
\hline 
\hline
\\
BCG Vaccine & 0.01 $\pm$ 0.04 & 0.00 $\pm$ 0.04 \\ 
Temperature & {\bf -0.29 $\pm$ 0.14} & -0.19 $\pm$ 0.13 \\ 
Relative Humidity & -0.10 $\pm$ 0.08 & -0.09 $\pm$ 0.09 \\ 
Tests / 1k & -0.11 $\pm$ 0.12 & -0.14 $\pm$ 0.11 \\ 
Positive Rate & {\bf 0.25 $\pm$ 0.08} & {\bf 0.27 $\pm$ 0.08} \\ 
UV Index & -0.01 $\pm$ 0.95 & -0.47 $\pm$ 0.89 \\ 

\\
\hline
\hline
\end{tabular}
   \caption{Comparison of univariate and multivariate fits to the data. Parameters whose mean are  more than 2$\sigma$ away from zero (temperature and positive rate) are shown in bold. Only Postive Rate is nonzero at more than $3\sigma$.}
   \label{tab:uni_multi}

\end{table}

\subsubsection{Varying Hyperpriors}

Since our goal in this analysis is to assess whether there is evidence for external factors such as climate, blood type etc... in addition to known country-specific factors, an important internal check is to look for potential sensitivity to our priors and hyperpriors.

In our hierarchical analysis, data for each country tries to pull the measured growth rates of countries to their own best-fit values, while the hierarchical nature of our model described in section (\ref{sec:probsample}) attempts to pull them all together. Between this tug of war, the algorithm searches for joint values of $\mathbf{\Theta}$ that will improve the fits to all the data. One concern might be that if we make the hyperpriors on the parent distributions much stronger or much weaker, we allow the algorithm to given less or more freedom to the base growth rates $G_{0c}$ which in turn may affect the best-fit parameters $\mathbf{\Theta}$ and our main conclusions. 

To test this we tightened the following hyperpriors and priors listed in section (\ref{sec:probsample}) by an order of magnitude to: 
\begin{itemize}
    \item $\mu_{G_0} \sim \mathcal{N}(20, 2^2)$; the mean (in percent) on the prior for the growth rate of each country, $G_{0c}$.
    \item $\sigma_{G_0} \sim \text{HalfNormal}(2)$; the standard deviation on the prior for the growth rate of each country.
    \item $\mathbf{\Theta} \sim \mathcal{N}(0, 0.1)$; the prior on the vector of parameters of key interest.
\end{itemize}

\begin{figure*}[!ht]

    \centerline{
    \includegraphics[width=1.2\linewidth]{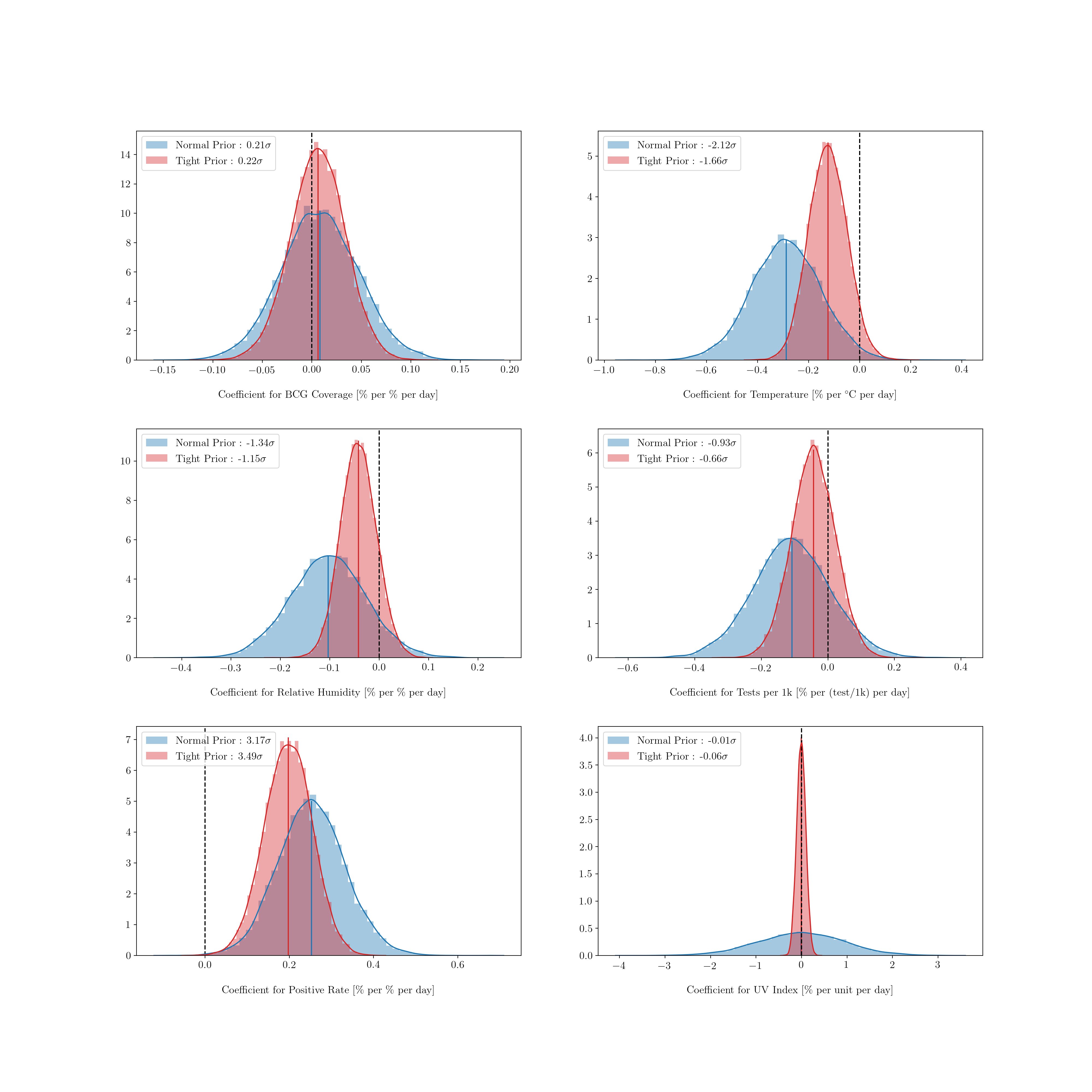}}
    \caption{Marginal distributions for the various factors in the case of the standard priors/hyperpriors and in the case where all the priors are tightened by a factor of ten. We see that the significance of the best-fit changes by less than $0.5\sigma$ although the means are typically shifted towards zero, other than BCG and Positive Rate. Our main conclusions are unchanged and stable to changing the priors dramatically. }
    \label{fig:marginals_all_tight}
\end{figure*}

The results of these changes are shown in Fig. (\ref{fig:marginals_all_tight}). As expected tightening the prior on the  $\mathbf{\Theta}$ pulled most parameters slightly closer to zero but also tighten the posterior so that none of our conclusions were altered: the statistical significance of all variables was unaltered.  

\subsection{Machine Learning Analysis}\label{sec:ml}

To provide a largely independent test of our Bayesian results we also undertook a machine learning analysis of the deaths data using random forest regression \cite{scikit}. Random forests are a powerful ensemble method that naturally provide feature selection capability, and are hundreds of times faster to run than our computationally intensive hierarchical Bayesian framework. While random forest does not provide estimates of statistical significance of factors it does allow us to rank additional factors in terms of importance and hence to explore the potential of additional explanatory features for inclusion in the main set of parameters, $\mathbf{\Theta}$, for the Bayesian analysis.   

To undertake the random forest regression we used the results from our No-Factor Bayesian run (i.e. $\mathbf{\Theta} = 0$) to obtain the base growth rates $G_{0c}$ for each country and used these as the target for the random forest regression with the $\mathbf{X_c}$ data as features. In the random forest analysis we augmented the data both with \pmt\ pollution and extra blood types data (namely A–, AB+, O–, AB–, B–).

Feature selection was done using the standard Random Forest impurity method\cite{scikit} and by ranking variables by the impact they had on the average Root-Mean-Square-Error (RMSE) of the regression: leaving out important explanatory variables is expected to significantly degrade the performance of the algorithm, leaving out irrelevant features does not. All feature importance scores were averages over 500 random 70-30 training-test splits of the data.  We found that the only variables that made more than one percent difference to the RMSE value were the Postive Rate (3.98\%) and A+ (2.28\%). This lead us to select A+ blood type as a variable in our separate full Bayesian analysis shown in section (\ref{sec:blood}). 

\begin{figure}[!ht]
    \includegraphics[width=0.45\textwidth]{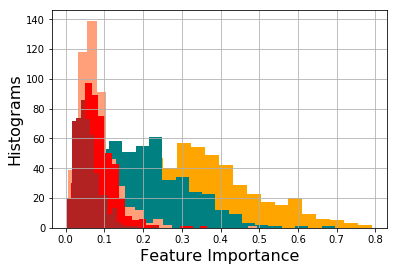}
    \caption{Histograms of feature importances from 500 random forest runs showing, from most important to least: Positive-Fraction (beige), A+ blood type (teal), BCG (salmon), Tests per 1k (red) and \pmt\ (dark brown).}
    \label{fig:BCG_countries}
\end{figure}

The results from the impurity-based feature selection are shown in Table (\ref{allresults_rf}). Again Positive Rate and A+ were the most significant features, followed by Tests per 1k and BCG vaccine coverage, \pmt\ and the other blood types at  significantly less importance. As a result blood types, other than A+, were not included in the Bayesian analysis since additional parameters significantly increased the computational complexity of the analysis, both because of increased time to convergence and an increase in the number of models to run (since we fit both multivariate and univariate models in all cases). The distribution of a selection of feature importances is shown in Fig. (\ref{fig:BCG_countries})

In summary the machine learning analysis confirms, to the extent that they overlap, the results of our much more intensive Bayesian analysis: Positive Rate is the most important feature, followed by A+ (subject to the caveats discussed in section (\ref{sec:blood}), while BCG is not relevant. 

\begin{table}
\centering
\begin{tabular}{lrr}
\hline
\hline
Feature &   Mean & Std. Dev \\
\hline
\hline
\\

Pos-Rate    &   0.36 &  0.14 \\
A+          &   0.20 &  0.11 \\
Tests/1k &  0.08 &  0.04 \\
BCG  &   0.08 &  0.05 \\
B+          &   0.08 &  0.06 \\
\pmt       &  0.06 &  0.03 \\
A--          &   0.04 &  0.03 \\
AB+         &   0.03 &  0.02 \\
B--          &   0.02 &  0.01 \\
AB--         &   0.02 &  0.02 \\
O--          &   0.02 &  0.02 \\
\\
\hline
\hline
\end{tabular}
	
\caption{Mean and standard deviation of impurity-based random forest feature importances from 500 runs with 100 estimators and a $30-70\%$ test-training  split. The sample density of feature importances is shown in Fig. (\ref{fig:BCG_countries}). Positive Fraction is selected as the most important feature, as with the Bayesian analysis. Climatic variables (temp, humidity and UV Index) were not included in the analysis.}
\label{allresults_rf}
\end{table}

\begin{table*}[!ht]
\centering
\begin{tabular}{lccccccc}
\hline
\hline
Country/Region & BCG (\%) & T ($^\circ$C) & RH (\%) & Tests/1k & Pos-Rate (\%) & A+ (\%) & UV \\
\hline
\hline
\\
Argentina & 58.5 & 21.6 & 51.0 & 0.8 & 12.7 & 34.3 & 1.3 \\ 
Australia / NSW & 38.9 & 19.7 & 74.2 & 17.3 & 3.4 & 31.0 & 1.4 \\ 
Austria & 53.1 & 1.3 & 60.3 & 21.5 & 15.0 & 33.0 & 1.0 \\ 
Canada / Alberta & 0.0 & -3.8 & 74.5 & 15.1 & 7.2 & 36.0 & 0.7 \\ 
Canada / BC & 0.0 & 7.9 & 70.9 & 15.1 & 4.4 & 36.0 & 0.8 \\ 
Canada / Ontario & 0.0 & -2.9 & 72.5 & 15.1 & 10.5 & 36.0 & 0.8 \\
Canada / Quebec & 0.0 & -10.8 & 68.8 & 14.8 & 27.3 & 36.0 & 0.7 \\
Chile & 93.1 & 15.8 & 62.8 & 6.6 & 13.7 & 8.7 & 1.5 \\ 
Colombia & 86.4 & 14.2 & 90.5 & 1.2 & 12.9 & 26.1 & 2.4 \\ 
Cuba & 46.9 & 22.3 & 79.3 & 2.6 & 7.8 & 32.8 & 2.5 \\ 
Czechia & 73.5 & 3.0 & 56.7 & 16.8 & 6.7 & 36.0 & 1.0 \\ 
Denmark & 50.7 & 4.8 & 61.6 & 17.3 & 16.1 & 37.0 & 0.7 \\ 
Estonia & 27.2 & 1.9 & 65.1 & 32.5 & 5.1 & 30.8 & 0.6 \\ 
Finland & 80.4 & 0.2 & 68.1 & 11.0 & 10.4 & 38.0 & 0.4 \\ 
France & 71.4 & 8.6 & 71.6 & 1.5 & 53.3 & 37.0 & 0.7 \\ 
Germany & 49.7 & 0.8 & 48.0 & 13.8 & 12.0 & 37.0 & 0.8 \\ 
Greece & 13.9 & 7.6 & 80.1 & 4.8 & 8.8 & 32.9 & 1.1 \\ 
Hungary & 82.3 & 6.2 & 57.4 & 5.2 & 8.9 & 33.0 & 0.9 \\ 
India & 96.9 & 30.1 & 29.3 & 0.3 & 14.5 & 20.8 & 2.4 \\ 
Israel & 29.0 & 15.3 & 58.3 & 28.4 & 10.4 & 34.0 & 1.7 \\ 
Italy & 0.0 & 4.5 & 46.5 & 1.2 & 40.0 & 36.0 & 1.2 \\ 
Japan & 70.0 & 4.4 & 64.8 & 1.4 & 16.8 & 39.8 & 1.2 \\  
Lithuania & 28.2 & 2.7 & 54.9 & 24.8 & 4.5 & 33.0 & 0.7 \\ 
Luxembourg & 0.0 & 6.7 & 55.5 & 57.9 & 12.5 & 37.0 & 1.0 \\ 
Mexico & 98.9 & 20.9 & 29.7 & 0.3 & 57.0 & 29.9 & 2.8 \\ 
Netherlands & 0.0 & 4.5 & 44.9 & 4.0 & 29.9 & 35.0 & 0.8 \\ 
Norway & 79.8 & -1.8 & 63.0 & 26.7 & 7.3 & 42.5 & 0.6 \\ 
Pakistan & 77.8 & 16.5 & 63.1 & 0.5 & 22.2 & 20.6 & 1.6 \\ 
Peru & 96.5 & 24.5 & 86.5 & 4.7 & 35.5 & 18.4 & 1.9 \\ 
Philippines & 61.8 & 27.5 & 82.5 & 0.5 & 14.7 & 28.9 & 2.4 \\ 
Poland & 80.8 & 4.3 & 51.1 & 5.6 & 11.6 & 31.3 & 0.8 \\ 
Slovenia & 75.2 & 6.3 & 47.1 & 20.6 & 4.4 & 33.0 & 1.1 \\ 
South Africa & 79.2 & 21.2 & 43.3 & 2.2 & 4.5 & 32.0 & 1.9 \\
South Korea & 51.3 & 8.2 & 50.7 & 11.1 & 3.3 & 32.8 & 1.3 \\ 
Sweden & 40.9 & -0.1 & 62.4 & 7.7 & 25.8 & 37.0 & 0.3 \\ 
Switzerland & 31.9 & -1.0 & 80.3 & 20.6 & 18.1 & 37.0 & 0.8 \\ 
Thailand & 53.4 & 31.1 & 60.8 & 0.6 & 10.0 & 16.9 & 2.4 \\ 
Turkey & 92.8 & 5.9 & 57.9 & 3.3 & 40.3 & 37.8 & 1.2 \\ 
US & 0.0 & 11.8 & 85.0 & 0.9 & 61.0 & 35.7 & 1.2 \\ 
United Kingdom & 67.2 & 3.5 & 71.3 & 1.6 & 19.6 & 35.0 & 0.5 \\ 
\\
\hline
\hline
\end{tabular}
\caption{Factor data, $\mathbf{X}_c$ for all regions and countries in our study. The columns are BCG fraction, temperature (T), relative humidity (RH), tests per 1000 population (Tests/1k), positive test rate (Pos-Rate), A+ blood type prevalence and UV Index (UV). We analyse blood type separately as discussed in section (\ref{sec:blood}). }
\label{allresults}
\end{table*}

\begin{table*}[!h]
    \scriptsize
    \def\arraystretch{1.4}
    \hskip-1.3cm\
    \begin{tabular}{lllllllllll} 
        \hline
        \hline
			Parameter 
		& \multicolumn{1}{p{1.0cm}}{\centering No Factors \\ (68\% CI)}
		& \multicolumn{1}{p{1.0cm}}{\centering BCG Vaccine \\ (68\% CI)}
		& \multicolumn{1}{p{1.0cm}}{\centering Temperature \\ (68\% CI)}
		& \multicolumn{1}{p{1.0cm}}{\centering Relative Humidity \\ (68\% CI)}
		& \multicolumn{1}{p{1.0cm}}{\centering Tests / 1k \\ (68\% CI)}
		& \multicolumn{1}{p{1.0cm}}{\centering Positive Rate \\ (68\% CI)}
		& \multicolumn{1}{p{1.0cm}}{\centering A+ Blood Type \\ (68\% CI)}
		& \multicolumn{1}{p{1.0cm}}{\centering UV Index \\ (68\% CI)}
		& \multicolumn{1}{p{1.0cm}}{\centering Excl. Tests \\ (68\% CI)}
		& \multicolumn{1}{p{1.0cm}}{\centering Incl. Tests \\ (68\% CI)} \\
		\hline \\[0.02em]

		$\mu_{D_0}$ & $5.60^{+0.40}_{-0.35}$ & $5.63^{+0.36}_{-0.38}$ & $5.61^{+0.39}_{-0.36}$ & $5.58^{+0.42}_{-0.32}$ & $5.72^{+0.28}_{-0.47}$ & $5.66^{+0.33}_{-0.40}$ & $5.56^{+0.43}_{-0.31}$ & $5.66^{+0.35}_{-0.40}$ & $5.74^{+0.34}_{-0.38}$ & $5.71\pm 0.37$ \\ 
		$\sigma_{D_0}$ & $2.18^{+0.28}_{-0.30}$ & $2.17^{+0.30}_{-0.29}$ & $2.15^{+0.29}_{-0.28}$ & $2.11^{+0.34}_{-0.24}$ & $2.18^{+0.29}_{-0.30}$ & $2.20^{+0.26}_{-0.31}$ & $2.22^{+0.23}_{-0.34}$ & $2.14^{+0.31}_{-0.26}$ & $2.16^{+0.28}_{-0.30}$ & $2.07^{+0.37}_{-0.21}$ \\ 
		$\mu_{G_0}$ & $20.6^{+1.2}_{-1.5}$ & $20.6^{+2.3}_{-2.6}$ & $22.3^{+1.7}_{-1.9}$ & $26.1^{+6.0}_{-5.8}$ & $22.3^{+1.5}_{-2.0}$ & $15.8^{+1.6}_{-2.0}$ & $17.0^{+6.9}_{-5.5}$ & $21.3^{+1.5}_{-1.9}$ & $26.1^{+7.0}_{-5.8}$ & $26.5^{+5.7}_{-7.1}$ \\ 
		$\sigma_{G_0}$ & $7.87^{+1.08}_{-0.90}$ & $7.98^{+1.13}_{-0.91}$ & $8.04^{+0.79}_{-1.17}$ & $7.78^{+1.18}_{-0.83}$ & $7.73^{+1.15}_{-0.84}$ & $6.70^{+1.09}_{-0.69}$ & $8.03^{+0.99}_{-1.04}$ & $7.71^{+1.19}_{-0.73}$ & $7.91^{+0.91}_{-1.09}$ & $6.28^{+1.06}_{-0.67}$ \\ 
		
		& & & & & & & & & & \\
 \hline
		 & & & & & & & & & & \\
		
        $\mathbf{G_0}_{\text{Argentina}}$ & $20.3^{+2.9}_{-2.6}$ & $20.2^{+3.9}_{-3.4}$ & $24.0^{+4.2}_{-3.5}$ & $24.9^{+6.2}_{-4.9}$ & $20.6^{+2.9}_{-2.7}$ & $16.5^{+2.9}_{-3.1}$ & $18.3^{+6.1}_{-7.7}$ & $21.4^{+2.8}_{-3.3}$ & $19.2^{+6.8}_{-6.7}$ & $27.8^{+5.4}_{-5.9}$ \\ 
		$\mathbf{G_0}_{\text{Australia / NSW}}$ & $5.5^{+2.3}_{-2.7}$ & $5.3\pm 3.0$ & $8.8^{+3.8}_{-3.2}$ & $11.6^{+7.9}_{-6.4}$ & $8.2^{+2.7}_{-3.6}$ & $4.5\pm 2.5$ & $1.9^{+7.2}_{-5.5}$ & $6.2^{+2.6}_{-3.1}$ & $19.3^{+7.4}_{-6.9}$ & $15.5^{+6.6}_{-6.7}$ \\ 
		$\mathbf{G_0}_{\text{Austria}}$ & $30.2\pm 2.6$ & $30.8^{+2.8}_{-4.0}$ & $30.8^{+2.4}_{-2.9}$ & $35.1^{+6.8}_{-5.4}$ & $32.7^{+3.8}_{-3.2}$ & $25.5^{+3.1}_{-2.5}$ & $29.0^{+5.0}_{-8.6}$ & $31.1^{+2.3}_{-3.3}$ & $33.5^{+6.7}_{-7.0}$ & $25.3^{+6.4}_{-5.5}$ \\ 
		$\mathbf{G_0}_{\text{Canada / Alberta}}$ & $13.4^{+1.5}_{-1.1}$ & $13.6\pm 1.3$ & $12.9\pm 1.4$ & $20.1^{+7.5}_{-6.4}$ & $16.0^{+1.7}_{-2.4}$ & $11.1^{+1.6}_{-1.3}$ & $10.1^{+7.2}_{-6.5}$ & $14.1^{+1.2}_{-1.6}$ & $31.3^{+6.6}_{-9.0}$ & $29.3^{+6.1}_{-7.5}$ \\ 
		$\mathbf{G_0}_{\text{Canada / BC}}$ & $10.4^{+2.2}_{-2.1}$ & $10.6^{+2.0}_{-2.3}$ & $12.1^{+2.5}_{-2.3}$ & $16.2^{+7.9}_{-5.8}$ & $12.4^{+3.0}_{-2.4}$ & $9.1^{+2.2}_{-2.1}$ & $7.8^{+6.6}_{-7.3}$ & $10.8^{+2.5}_{-2.1}$ & $34.1^{+9.0}_{-9.4}$ & $24.0^{+6.1}_{-6.9}$ \\ 
		$\mathbf{G_0}_{\text{Canada / Ontario}}$ & $24.8^{+2.0}_{-1.1}$ & $25.4^{+1.5}_{-1.6}$ & $24.8^{+1.5}_{-1.7}$ & $32.1^{+6.7}_{-6.9}$ & $27.0^{+2.7}_{-1.9}$ & $22.3^{+1.5}_{-2.0}$ & $21.8^{+7.5}_{-6.4}$ & $25.9^{+1.5}_{-1.9}$ & $19.6^{+6.5}_{-8.4}$ & $27.4^{+8.7}_{-7.1}$ \\ 
		$\mathbf{G_0}_{\text{Canada / Quebec}}$ & $24.5\pm 1.1$ & $24.3^{+1.2}_{-1.0}$ & $22.1^{+2.1}_{-1.4}$ & $29.7^{+7.4}_{-5.3}$ & $26.8^{+1.6}_{-2.3}$ & $17.2^{+2.2}_{-2.6}$ & $21.0^{+7.1}_{-6.6}$ & $24.8^{+1.3}_{-1.2}$ & $35.6^{+5.8}_{-7.5}$ & $26.8^{+5.7}_{-7.7}$ \\ 
		$\mathbf{G_0}_{\text{Chile}}$ & $20.1^{+2.1}_{-1.7}$ & $20.2^{+4.3}_{-4.1}$ & $22.5^{+3.5}_{-2.0}$ & $25.0^{+7.3}_{-5.0}$ & $21.3^{+1.9}_{-2.2}$ & $16.2\pm 2.3$ & $18.6^{+3.5}_{-1.9}$ & $20.4^{+3.0}_{-1.6}$ & $17.3^{+6.4}_{-7.6}$ & $20.6^{+7.3}_{-7.0}$ \\ 
		$\mathbf{G_0}_{\text{Colombia}}$ & $19.2^{+1.2}_{-1.4}$ & $19.3^{+3.5}_{-3.9}$ & $22.4^{+1.7}_{-2.8}$ & $27.0^{+8.7}_{-7.9}$ & $19.5^{+1.1}_{-1.5}$ & $15.0^{+2.1}_{-1.3}$ & $16.5^{+5.5}_{-4.7}$ & $20.1^{+2.8}_{-2.2}$ & $18.5^{+6.4}_{-6.9}$ & $28.2^{+6.6}_{-7.8}$ \\ 
		$\mathbf{G_0}_{\text{Cuba}}$ & $11.3^{+1.5}_{-1.7}$ & $11.4^{+2.3}_{-2.7}$ & $15.5\pm 3.3$ & $19.1^{+6.7}_{-7.9}$ & $11.9^{+1.4}_{-1.9}$ & $8.6^{+2.2}_{-1.3}$ & $8.6^{+6.0}_{-6.6}$ & $12.9^{+2.3}_{-3.2}$ & $27.2^{+5.8}_{-5.3}$ & $31.0^{+7.7}_{-6.2}$ \\ 
		$\mathbf{G_0}_{\text{Czechia}}$ & $26.3^{+3.2}_{-2.2}$ & $26.1^{+4.7}_{-3.2}$ & $27.6^{+2.7}_{-2.9}$ & $32.9^{+5.0}_{-6.8}$ & $28.8^{+3.7}_{-2.9}$ & $23.9^{+3.2}_{-2.3}$ & $24.2^{+6.8}_{-7.8}$ & $27.6^{+2.6}_{-3.3}$ & $28.7^{+7.6}_{-9.6}$ & $24.5^{+6.7}_{-8.2}$ \\ 
		$\mathbf{G_0}_{\text{Denmark}}$ & $26.4^{+2.8}_{-1.9}$ & $26.7^{+3.2}_{-3.0}$ & $27.8^{+2.5}_{-2.4}$ & $32.9^{+5.9}_{-6.4}$ & $29.4^{+2.8}_{-3.2}$ & $22.2^{+2.7}_{-2.6}$ & $23.9^{+7.0}_{-7.6}$ & $27.7^{+2.0}_{-2.9}$ & $34.5\pm 7.4$ & $21.1^{+6.1}_{-6.9}$ \\ 
		$\mathbf{G_0}_{\text{Estonia}}$ & $14.2^{+2.4}_{-2.0}$ & $14.4^{+2.6}_{-2.4}$ & $14.8^{+2.5}_{-2.0}$ & $20.0^{+6.8}_{-5.8}$ & $19.2^{+3.5}_{-4.4}$ & $12.9\pm 2.2$ & $11.7^{+6.5}_{-6.0}$ & $14.6^{+2.5}_{-2.1}$ & $27.8^{+5.9}_{-6.1}$ & $19.2^{+5.1}_{-7.2}$ \\ 
		$\mathbf{G_0}_{\text{Finland}}$ & $14.1^{+1.6}_{-1.4}$ & $14.1^{+3.7}_{-3.5}$ & $14.2^{+1.6}_{-1.3}$ & $20.8^{+6.1}_{-6.7}$ & $15.8^{+2.0}_{-1.8}$ & $11.1^{+2.0}_{-1.4}$ & $10.5^{+7.8}_{-6.8}$ & $14.6^{+1.4}_{-1.7}$ & $24.8^{+6.1}_{-5.9}$ & $21.9^{+5.0}_{-7.1}$ \\ 
		$\mathbf{G_0}_{\text{France}}$ & $28.4^{+1.0}_{-1.8}$ & $28.1^{+3.0}_{-3.4}$ & $29.8^{+1.7}_{-1.9}$ & $34.6^{+6.7}_{-6.6}$ & $28.3\pm 1.4$ & $14.0^{+4.1}_{-4.6}$ & $25.3^{+6.8}_{-7.5}$ & $28.4^{+1.5}_{-1.6}$ & $23.7^{+6.0}_{-5.9}$ & $29.8^{+7.5}_{-9.6}$ \\ 
		$\mathbf{G_0}_{\text{Germany}}$ & $30.4^{+1.1}_{-0.9}$ & $30.4^{+2.3}_{-2.2}$ & $30.6^{+1.1}_{-1.0}$ & $34.7^{+4.7}_{-4.3}$ & $32.2^{+2.0}_{-1.6}$ & $27.0^{+1.5}_{-1.3}$ & $26.7^{+7.5}_{-6.5}$ & $30.9^{+1.2}_{-1.3}$ & $20.4^{+5.5}_{-6.0}$ & $23.6^{+6.8}_{-8.7}$ \\ 
		$\mathbf{G_0}_{\text{Greece}}$ & $17.3\pm 1.7$ & $17.4^{+1.6}_{-2.0}$ & $18.9^{+1.8}_{-2.1}$ & $24.1^{+8.1}_{-6.9}$ & $18.2^{+1.5}_{-2.0}$ & $14.5^{+2.2}_{-1.4}$ & $15.8^{+4.8}_{-8.0}$ & $18.2^{+1.6}_{-2.3}$ & $13.9^{+6.4}_{-4.4}$ & $22.1^{+6.7}_{-6.2}$ \\ 
		$\mathbf{G_0}_{\text{Hungary}}$ & $13.0^{+2.1}_{-1.6}$ & $13.2^{+3.7}_{-3.8}$ & $14.7^{+1.8}_{-2.3}$ & $18.2^{+6.0}_{-5.1}$ & $14.2^{+1.7}_{-2.1}$ & $11.0^{+1.7}_{-2.2}$ & $9.4^{+7.5}_{-5.5}$ & $13.6^{+2.1}_{-1.9}$ & $19.8^{+6.7}_{-6.2}$ & $31.6^{+8.6}_{-6.7}$ \\ 
		$\mathbf{G_0}_{\text{India}}$ & $20.1^{+0.7}_{-0.7}$ & $19.8^{+4.1}_{-3.8}$ & $26.1^{+3.7}_{-4.2}$ & $23.1^{+2.5}_{-3.1}$ & $20.2^{+0.6}_{-0.7}$ & $16.0^{+1.3}_{-1.4}$ & $18.5^{+3.8}_{-4.2}$ & $21.4^{+2.1}_{-2.4}$ & $30.7^{+6.1}_{-5.0}$ & $19.0^{+6.4}_{-6.5}$ \\ 
		$\mathbf{G_0}_{\text{Israel}}$ & $19.8^{+1.8}_{-1.7}$ & $19.7^{+2.3}_{-1.9}$ & $22.6^{+2.9}_{-2.4}$ & $25.4^{+5.6}_{-5.7}$ & $23.5^{+3.9}_{-3.1}$ & $16.8^{+2.2}_{-1.7}$ & $17.8^{+5.8}_{-7.3}$ & $20.7^{+2.4}_{-2.3}$ & $27.8^{+6.5}_{-6.4}$ & $29.9^{+9.1}_{-9.6}$ \\ 
		$\mathbf{G_0}_{\text{Italy}}$ & $35.1^{+0.7}_{-0.8}$ & $35.2^{+0.6}_{-0.9}$ & $35.9^{+1.0}_{-0.9}$ & $39.8^{+3.8}_{-4.8}$ & $35.1^{+0.9}_{-0.7}$ & $25.2^{+2.9}_{-3.1}$ & $31.3^{+7.4}_{-6.1}$ & $35.9^{+0.9}_{-1.6}$ & $45.7^{+7.6}_{-7.2}$ & $25.8^{+6.7}_{-9.5}$ \\ 
		$\mathbf{G_0}_{\text{Japan}}$ & $11.3^{+1.9}_{-1.5}$ & $11.4^{+3.4}_{-3.2}$ & $12.9^{+1.4}_{-2.3}$ & $18.3^{+5.4}_{-7.0}$ & $12.0^{+1.6}_{-1.9}$ & $6.9^{+2.1}_{-2.5}$ & $8.5^{+7.3}_{-8.0}$ & $12.1^{+2.2}_{-1.9}$ & $27.4^{+7.5}_{-6.9}$ & $44.1^{+6.1}_{-8.1}$ \\  
		$\mathbf{G_0}_{\text{Lithuania}}$ & $10.5^{+1.2}_{-1.9}$ & $10.2^{+1.7}_{-2.2}$ & $10.8^{+1.5}_{-1.8}$ & $15.4^{+5.2}_{-5.5}$ & $13.2^{+3.5}_{-2.6}$ & $8.9^{+1.6}_{-1.5}$ & $8.3^{+5.4}_{-7.4}$ & $10.6^{+1.5}_{-1.9}$ & $21.8^{+5.3}_{-5.8}$ & $31.9^{+4.7}_{-5.5}$ \\ 
		$\mathbf{G_0}_{\text{Luxembourg}}$ & $15.6^{+2.0}_{-2.2}$ & $15.4\pm 2.2$ & $16.8^{+2.3}_{-2.4}$ & $21.1^{+5.2}_{-6.1}$ & $24.4^{+4.8}_{-7.7}$ & $12.2^{+2.2}_{-2.5}$ & $12.4^{+7.1}_{-7.4}$ & $15.8^{+2.4}_{-2.3}$ & $28.2^{+9.3}_{-9.5}$ & $28.8^{+5.3}_{-7.2}$ \\ 
		$\mathbf{G_0}_{\text{Mexico}}$ & $26.2^{+1.4}_{-1.6}$ & $26.5^{+3.8}_{-4.7}$ & $30.0^{+3.5}_{-2.9}$ & $29.4^{+3.9}_{-3.7}$ & $26.2\pm 1.5$ & $12.3^{+4.7}_{-3.7}$ & $22.9^{+6.6}_{-5.1}$ & $28.0^{+2.1}_{-3.3}$ & $21.6^{+8.5}_{-7.8}$ & $32.5^{+7.3}_{-6.7}$ \\ 
		$\mathbf{G_0}_{\text{Netherlands}}$ & $34.4^{+1.7}_{-1.3}$ & $34.7\pm 1.5$ & $35.4^{+1.8}_{-1.4}$ & $39.2^{+4.2}_{-4.8}$ & $35.5^{+1.3}_{-1.8}$ & $26.3^{+3.1}_{-2.4}$ & $32.6^{+6.0}_{-7.5}$ & $35.3^{+1.4}_{-1.8}$ & $23.9^{+5.2}_{-7.0}$ & $22.5^{+6.4}_{-5.4}$ \\ 
		$\mathbf{G_0}_{\text{Norway}}$ & $16.2^{+1.2}_{-1.0}$ & $15.9^{+3.6}_{-3.2}$ & $16.0^{+1.0}_{-1.2}$ & $22.0^{+5.9}_{-5.8}$ & $20.0^{+3.0}_{-3.1}$ & $14.4^{+1.0}_{-1.4}$ & $13.2^{+7.5}_{-8.4}$ & $16.7^{+0.9}_{-1.4}$ & $31.2^{+7.6}_{-5.4}$ & $35.3^{+6.4}_{-8.0}$ \\ 
		$\mathbf{G_0}_{\text{Pakistan}}$ & $19.2^{+2.1}_{-2.3}$ & $18.7^{+4.3}_{-3.3}$ & $22.8^{+2.7}_{-3.5}$ & $24.2^{+7.0}_{-5.3}$ & $19.1^{+2.6}_{-1.9}$ & $13.1^{+3.0}_{-2.6}$ & $17.4^{+4.7}_{-4.6}$ & $19.2^{+3.4}_{-1.9}$ & $14.4^{+8.2}_{-7.0}$ & $30.2^{+7.3}_{-7.0}$ \\ 
		$\mathbf{G_0}_{\text{Peru}}$ & $23.1^{+1.6}_{-1.5}$ & $22.8^{+4.4}_{-3.9}$ & $27.3^{+4.0}_{-3.0}$ & $28.5^{+10.7}_{-5.3}$ & $24.2^{+1.3}_{-2.0}$ & $12.6^{+3.2}_{-3.8}$ & $21.1^{+4.4}_{-3.5}$ & $24.2^{+2.1}_{-2.5}$ & $31.8^{+6.4}_{-7.4}$ & $18.7^{+7.8}_{-7.4}$ \\ 
		$\mathbf{G_0}_{\text{Philippines}}$ & $15.9^{+1.3}_{-1.0}$ & $15.9^{+2.8}_{-2.7}$ & $20.8^{+4.1}_{-3.3}$ & $24.6^{+6.5}_{-8.7}$ & $15.8^{+1.4}_{-0.9}$ & $11.8^{+1.7}_{-1.6}$ & $13.4^{+5.4}_{-5.7}$ & $17.5^{+2.1}_{-2.9}$ & $39.9^{+4.4}_{-4.3}$ & $26.5^{+8.4}_{-9.2}$ \\ 
		$\mathbf{G_0}_{\text{Poland}}$ & $19.3^{+1.0}_{-0.8}$ & $19.4^{+3.2}_{-3.6}$ & $20.2^{+1.0}_{-1.1}$ & $24.6^{+4.2}_{-5.3}$ & $20.1\pm 1.1$ & $16.1\pm 1.3$ & $17.1^{+5.7}_{-6.2}$ & $19.7^{+1.1}_{-1.2}$ & $29.3^{+5.3}_{-7.5}$ & $25.5^{+5.4}_{-9.3}$ \\ 
		$\mathbf{G_0}_{\text{Slovenia}}$ & $14.8^{+1.7}_{-1.4}$ & $14.8^{+3.5}_{-3.3}$ & $16.7^{+1.3}_{-2.2}$ & $19.0^{+4.9}_{-4.3}$ & $17.5^{+3.1}_{-2.4}$ & $13.7^{+1.6}_{-1.7}$ & $12.0^{+6.5}_{-6.3}$ & $15.0^{+2.3}_{-1.5}$ & $39.2^{+4.7}_{-3.9}$ & $33.6^{+5.0}_{-4.6}$ \\ 
		$\mathbf{G_0}_{\text{South Africa}}$ & $12.9^{+1.8}_{-1.3}$ & $13.3^{+3.3}_{-3.8}$ & $17.4^{+2.9}_{-3.4}$ & $17.0^{+4.8}_{-3.9}$ & $13.6^{+1.4}_{-1.7}$ & $11.7\pm 1.6$ & $11.0^{+5.7}_{-6.7}$ & $14.1\pm 2.3$ & $25.9^{+6.9}_{-8.0}$ & $18.9^{+7.3}_{-9.4}$ \\ 
		
		$\mathbf{G_0}_{\text{South Korea}}$ & $18.4^{+2.5}_{-2.6}$ & $17.9^{+3.8}_{-2.9}$ & $20.2^{+2.6}_{-3.0}$ & $24.6^{+3.9}_{-6.9}$ & $20.0^{+2.8}_{-2.9}$ & $16.8^{+2.9}_{-2.3}$ & $16.2^{+6.2}_{-7.2}$ & $19.6^{+2.2}_{-3.3}$ & $17.5^{+7.1}_{-6.2}$ & $34.0^{+5.1}_{-5.7}$ \\
		
		$\mathbf{G_0}_{\text{Sweden}}$ & $20.4^{+0.3}_{-0.4}$ & $20.3^{+1.8}_{-1.6}$ & $20.4^{+0.3}_{-0.4}$ & $25.4^{+6.5}_{-4.9}$ & $21.5\pm 0.9$ & $13.4\pm 2.1$ & $16.7^{+7.5}_{-6.4}$ & $20.6^{+0.4}_{-0.5}$ & $27.5^{+7.0}_{-7.7}$ & $21.1^{+5.4}_{-7.2}$ \\ 
		$\mathbf{G_0}_{\text{Switzerland}}$ & $27.0^{+1.4}_{-1.3}$ & $27.0\pm 1.9$ & $27.0^{+1.1}_{-1.5}$ & $33.8^{+8.0}_{-6.7}$ & $29.6^{+2.8}_{-2.4}$ & $22.4^{+1.6}_{-2.2}$ & $23.3^{+7.6}_{-6.5}$ & $27.8^{+1.1}_{-1.9}$ & $33.3^{+5.7}_{-4.7}$ & $27.2^{+6.6}_{-8.1}$ \\ 
		$\mathbf{G_0}_{\text{Thailand}}$ & $16.9^{+2.3}_{-2.5}$ & $16.4^{+3.6}_{-2.8}$ & $22.9^{+4.2}_{-4.9}$ & $21.2^{+7.4}_{-4.7}$ & $17.0\pm 2.4$ & $14.1^{+2.4}_{-2.6}$ & $15.6^{+3.9}_{-4.4}$ & $19.2^{+1.9}_{-4.4}$ & $32.8^{+7.7}_{-7.5}$ & $23.9^{+7.6}_{-6.3}$ \\ 
		$\mathbf{G_0}_{\text{Turkey}}$ & $29.1^{+0.8}_{-1.1}$ & $28.9^{+3.8}_{-3.9}$ & $30.1\pm 1.2$ & $35.6^{+4.2}_{-6.6}$ & $29.4^{+1.1}_{-1.0}$ & $18.1^{+3.2}_{-3.4}$ & $26.0^{+6.9}_{-7.3}$ & $29.4^{+1.6}_{-1.3}$ & $19.7^{+5.0}_{-6.5}$ & $22.4^{+5.5}_{-5.7}$ \\ 
		$\mathbf{G_0}_{\text{US}}$ & $19.2^{+1.4}_{-1.8}$ & $18.7^{+1.8}_{-1.4}$ & $21.1^{+2.4}_{-2.1}$ & $26.9\pm 7.8$ & $19.1\pm 1.6$ & $2.5^{+5.2}_{-4.9}$ & $15.7^{+7.0}_{-6.7}$ & $19.5\pm 1.9$ & $30.7^{+6.8}_{-6.5}$ & $19.0^{+7.8}_{-6.7}$ \\ 
		$\mathbf{G_0}_{\text{United Kingdom}}$ & $41.0\pm 2.8$ & $40.9^{+4.0}_{-3.8}$ & $41.8^{+2.6}_{-3.0}$ & $47.5^{+7.1}_{-7.0}$ & $41.9^{+2.3}_{-3.4}$ & $34.6^{+2.5}_{-4.1}$ & $38.6^{+6.3}_{-8.1}$ & $41.3\pm 2.8$ & $27.0^{+6.8}_{-7.6}$ & $30.2^{+7.1}_{-5.6}$ \\ 
		
		& & & & & & & & & & \\
 \hline
		 & & & & & & & & & & \\

		$\mathbf{\Theta}_{\text{BCG}}$ & -- & $0.00^{+0.04}_{-0.04}$ & -- & -- & -- & -- & -- & -- & $0.02^{+0.04}_{-0.04}$ & $-0.01^{+0.05}_{-0.03}$ \\ 
		$\mathbf{\Theta}_{\text{Temp}}$ & -- & -- & $-0.21^{+0.15}_{-0.11}$ & -- & -- & -- & -- & -- & $-0.24^{+0.16}_{-0.15}$ & $-0.30^{+0.15}_{-0.12}$ \\ 
		$\mathbf{\Theta}_{\text{RH}}$ & -- & -- & -- & $-0.09^{+0.09}_{-0.10}$ & -- & -- & -- & -- & $-0.09^{+0.09}_{-0.09}$ & $-0.10^{+0.08}_{-0.08}$ \\ 
		$\mathbf{\Theta}_{\text{Tests/1k}}$ & -- & -- & -- & -- & $-0.12^{+0.08}_{-0.13}$ & -- & -- & -- & -- & $-0.11^{+0.12}_{-0.11}$ \\ 
		$\mathbf{\Theta}_{\text{Pos.Rate}}$ & -- & -- & -- & -- & -- & 
		$0.27^{+0.08}_{-0.08}$ & -- & -- & -- & $0.25^{+0.08}_{-0.08}$ \\ 
		$\mathbf{\Theta}_{\text{A+}}$ & -- & -- & -- & -- & -- & -- & $0.08^{+0.19}_{-0.18}$ & -- & -- & -- \\ 
		$\mathbf{\Theta}_{\text{UV}}$ & -- & -- & -- & -- & -- & -- & -- & $-0.53^{+0.94}_{-0.87}$ & $0.10^{+0.93}_{-1.01}$ & $-0.01^{+0.99}_{-0.93}$ \\   
		\\
				[0.02em]
				\hline 
		\hline
    \end{tabular}
    \caption{Base growth rates, $\mathbf{\Theta}$ parameters and hierarchical hyperparameters for the 40 regions. For space reasons we use the abbreviations RH (relative humidity), NSW (New South Wales) and  BC (British Columbia). }
    \label{tab:all_results}
\end{table*}






\begin{thebibliography}{}
\scriptsize
\setlength{\itemsep}{0pt}



\bibitem{imperial2}
S. Flaxman, S. Mishra, A. Gandy, \etal. Estimating the effects of non-pharmaceutical interventions on COVID-19 in Europe. Nature (2020). https://doi.org/10.1038/s41586-020-2405-7


\bibitem{gdp}
N. S. Diffenbaugh, M.  Burke,
Proceedings of the National Academy of Sciences May 2019, 116 (20) 9808-9813; DOI: 10.1073/pnas.1816020116.
https://www.pnas.org/content/116/20/9808

\bibitem{worldometer} https://www.worldometers.info/coronavirus/ downloaded 10 April 2020. 


\bibitem{strains}
B Korber, {\em et al}., bioRxiv 2020.04.29.069054; doi: https://doi.org/10.1101/2020.04.29.069054


\bibitem{luis}
A. Stier, M. Berman, and L. Bettencourt, 2020; 
https://arxiv.org/abs/2003.10376

\bibitem{bcg_data2}
https://www.who.int/data/gho/data/indicators/indicator-details/GHO/bcg-immunization-coverage-among-1-year-olds-(-)

\bibitem{age_data}
https://population.un.org/wpp/Download/Standard/Population/


\bibitem{weather1}
M.	Araújo and B. Naimi,
medRxiv 2020.03.12.20034728; doi: https://doi.org/10.1101/2020.03.12.20034728

\bibitem{weather2}
H. V. Fineberg {\em et al.}, Rapid Expert Consultation,  https://www.nap.edu/read/25771/chapter/1

\bibitem{weather3}
J. Wang {\em et al}., 
https://arxiv.org/abs/2003.05003
See also: http://covid19-report.com/\#/r-value 

\bibitem{weather4}
R. Baker {\em et al.}, 
medRxiv 2020.04.03.20052787; doi: https://doi.org/10.1101/2020.04.03.20052787

\bibitem{weather5}
P. Shi {\em et al},
medRxiv 2020.03.22.20038919; doi: https://doi.org/10.1101/2020.03.22.20038919

\bibitem{weather6}
M. Sajadi \etal,  https://ssrn.com/abstract=3550308; http://dx.doi.org/10.2139/ssrn.3550308

\bibitem{weather7}
J. Xi and Y. Zhu, 
Science of The Total Environment
724, 138201, 2020, 
https://doi.org/10.1016/j.scitotenv.2020.138201

\bibitem{weather8}
J. Ma \etal,
Science of The Total Environment, 
724, 138226, 2020;
https://doi.org/10.1016/j.scitotenv.2020.138226

\bibitem{weather9}
A. Anis, 
https://ssrn.com/abstract=3567639; http://dx.doi.org/10.2139/ssrn.3567639

\bibitem{weather10}
S. Pawar \etal, 
medRxiv 2020.03.29.20044461; doi: https://doi.org/10.1101/2020.03.29.20044461

\bibitem{weather11}
D. Gupta, 
http://dx.doi.org/10.2139/ssrn.3558470

\bibitem{weather12}
A. Notari, medRxiv 2020.03.26.20044529; doi: https://doi.org/10.1101/2020.03.26.20044529

\bibitem{blood1}
J. Zhao \etal, medRxiv 2020.03.11.20031096; doi: https://doi.org/10.1101/2020.03.11.20031096

\bibitem{blood2}
M. Zietz, N. P. Tatonetti
medRxiv 2020.04.08.20058073; doi: https://doi.org/10.1101/2020.04.08.20058073


\bibitem{bcg1}
A. Miller, \etal\
medRxiv 2020.03.24.20042937; doi: https://doi.org/10.1101/2020.03.24.20042937

\bibitem{bcg2}
L. E. Escobar, A. Molina-Cruz, C. Barillas-Mury
medRxiv 2020.05.05.20091975; doi: https://doi.org/10.1101/2020.05.05.20091975

\bibitem{bcg_neg}
S. Singh,  medRxiv 2020.04.11.20062232; doi: https://doi.org/10.1101/2020.04.11.20062232

\bibitem{bcg_neg2}
M. Asahara, medRxiv 2020.04.17.20068601; doi: https://doi.org/10.1101/2020.04.17.20068601

\bibitem{pollution}
E. Conticini, B. Frediani, D. Caro, 
Environmental Pollution, 2020; 114465 DOI: 10.1016/j.envpol.2020.114465

\bibitem{pollution2}
X. Wu, \etal,
medRxiv 2020.04.05.20054502; doi: https://doi.org/10.1101/2020.04.05.20054502

\bibitem{pollution3}
M. Travaglio, {\em et al},
medRxiv 2020.04.16.20067405; doi: https://doi.org/10.1101/2020.04.16.20067405

\bibitem{pollution4}
D. Liang, L. Shi, J. Zhao, \etal, 
Preprint. medRxiv. 2020;2020.05.04.20090746. doi:10.1101/2020.05.04.20090746

\bibitem{pollution5}
V. Bianconi \etal, 
Archives of Medical Science. 2020. doi:10.5114/aoms.2020.95336.

\bibitem{haplogroup}
A. Gómez-Carballa, \etal\,
bioRxiv 2020.05.19.097410; doi: https://doi.org/10.1101/2020.05.19.097410

\bibitem{bloodtype1}
https://www.rhesusnegative.net/themission/bloodtypefrequencies/

\bibitem{bloodtype2}
https://en.wikipedia.org/wiki/Blood\_type\_distribution\_by\_country

\bibitem{pm25}
https://data.worldbank.org/indicator/EN.ATM.PM25.MC.M3

\bibitem{vitd}
G. Davies, A. R Garami, J. C Byers, 
medRxiv 2020.05.01.20087965; doi: https://doi.org/10.1101/2020.05.01.20087965

\bibitem{vitd2}
D. De Smet, \etal 
medRxiv 2020.05.01.20079376; doi: https://doi.org/10.1101/2020.05.01.20079376

\bibitem{vitd3}
K. Razdan, K. Singh, D. Singh, 
Med Drug Discov. 2020;100051. doi:10.1016/j.medidd.2020.100051

\bibitem{uv_vitd}
O. Engelsen, 
Nutrients. 2010;2(5):482-495. doi:10.3390/nu2050482

\bibitem{gelman}
A. Gelman and J. Hill, 2006. Data analysis using regression and multilevel/hierarchical models. Cambridge university press.

\bibitem{bic}
R. E. Kass  and  A. E.   Raftery, Journal of the American Statistical Association, 90,
430, 773, 1995. DOI: 10.1080/01621459.1995.10476572

\bibitem{NUTS}
M. Hoffman, A. Gelman. The No-U-Turn sampler: adaptively setting path lengths in Hamiltonian Monte Carlo. J. Mach. Learn. Res.. 2014 Apr 1;15(1):1593-623. https://arxiv.org/abs/1111.4246

\bibitem{numpyro}
D. Phan, N. Pradhan and M. Jankowiak, arXiv:1912.11554;  https://github.com/pyro-ppl/numpyro

\bibitem{climate_data}
https://darksky.net

\bibitem{covid_data}
https://github.com/CSSEGISandData/COVID-19

\bibitem{bcg_data1}
http://www.bcgatlas.org/index.php

\bibitem{bcg_atlas}
A. Zwerling, M.A. Behr, {\em et al},
PLoS medicine, 8, 3, 2011. 

\bibitem{excess1}
C. Modi \etal,  
medRxiv 2020.04.15.20067074; doi: https://doi.org/10.1101/2020.04.15.20067074

\bibitem{excess2}
S. Vandoros,
Social Science \& Medicine doi: 10.1016/j.socscimed.2020.113101


\bibitem{falseneg1}
P. Wikramaratna, R. S Paton, M. Ghafari, J. Lourenco, 
medRxiv 2020.04.05.20053355; doi: https://doi.org/10.1101/2020.04.05.20053355

\bibitem{falseneg2}
I. Arevalo-Rodriguez {\em et al},
medRxiv 2020.04.16.20066787; doi: https://doi.org/10.1101/2020.04.16.20066787

\bibitem{chainconsumer}
S. Hinton, The Journal of Open Source Software, 1, 00045; 2016 10.21105/joss.00045

\bibitem{scikit}
F. Pedregosa, {\em et al}, Journal of Machine Learning Research, 12, 2825, 2011

\end{thebibliography}
\end{document}